%% file: draft_October2021.tex
\documentclass[a4paper, 11pt, numbers=noenddot, intlimits]{scrartcl}

\usepackage{lmodern}
\usepackage{hyperref}

\usepackage[utf8]{inputenc} 

\usepackage{float} 

\usepackage[sc]{mathpazo} 
\usepackage[T1]{fontenc}

\usepackage{amsmath,amsfonts,amssymb}
\newcommand{\qed}{\hfill $\Box$}
\usepackage{geometry} 
\usepackage{svg}
\usepackage{graphicx}
\usepackage{epstopdf}
\usepackage{adjustbox}

\usepackage{url}

\usepackage{booktabs} 
\usepackage{array} 
\usepackage{paralist} 
\usepackage{verbatim} 
\usepackage{subfig} 

\usepackage{fancyhdr} 
\usepackage{sectsty}
\allsectionsfont{\sffamily\mdseries\upshape} 

\usepackage[nottoc,notlof,notlot]{tocbibind} 
\usepackage[titles,subfigure]{tocloft} 

\newtheorem{prop}{Proposition}[section]

\newtheorem{theorem}[prop]{Theorem}

\usepackage{color}
\usepackage{colordvi}
\definecolor{magenta}{rgb}{.5,0,.5}
\definecolor{black}{rgb}{1.0,1.0,1.0}
\definecolor{magenta}{rgb}{.1,0,.3}
\definecolor{gruen}{rgb}{0.2,0.5,.5}
\definecolor{light}{rgb}{ 0.992, 0.961,  0.902}
\definecolor{Tan}{rgb}{ 0.992, 0.9,  0.902}

\definecolor{darkblue}{rgb}{0.2, 0.2, 0.6}

\hypersetup{
	colorlinks   = true,
	allcolors    = darkblue
}

\newcommand{\komment}[1]{{}}

\usepackage{titlesec}
\titleformat*{\section}{\Large\bfseries}
\titleformat*{\subsection}{\normalsize\itshape}

\titleformat*{\paragraph}{\bfseries}

\DeclareOldFontCommand{\bf}{\normalfont\bfseries}{\mathbf}



\deffootnote[1.25em]{1.25em}{1em}{\textsuperscript{\thefootnotemark}~}

\begin{document}

\title{A model of opinion dynamics with echo chambers explains the spatial distribution of vaccine hesitancy\footnote{J.~M\"uller is affiliated to the Technical University of Munich and the Helmholtzzentrum Munich (\href{mailto:johannes.mueller@mytum.de}{johannes.mueller@mytum.de}), A.~Tellier is affiliated to the Technical University of Munich (\href{mailto:tellier@wzw.tum.de}{tellier@wzw.tum.de}), and M.~Kurschilgen is affiliated to the Technical University of Munich, the Max Planck Institute for Research on Collective Goods, and the Stanford Graduate School of Business (\href{mailto:m.kurschilgen@tum.de}{m.kurschilgen@tum.de})}}
\author{Johannes M\"uller, Aur\'elien Tellier and Michael Kurschilgen}

\clearpage\maketitle
\thispagestyle{empty}

\begin{abstract} 
\noindent Vaccination hesitancy is a major obstacle to achieving and maintaining herd immunity. It is therefore of prime importance for public health authorities to understand the dynamics of an anti-vaccine opinion in the population. We introduce a novel mathematical model of opinion dynamics with spatial reinforcement, which can generate echo chambers, i.e. opinion bubbles in which information that is incompatible with one's entrenched worldview, is likely disregarded. In a first mathematical part, we scale the model both to a deterministic limit and to a weak-effects limit, and obtain bifurcations, phase transitions, and the invariant measure. In a second part, we fit our model to measles and meningococci vaccination coverage across 413 districts in Germany. We reveal that strong echo chambers explain the occurrence and persistence of the anti-vaccination opinion. We predict and compare the effectiveness of different policies aimed at influencing opinion dynamics in order to increase vaccination uptake. 
\end{abstract}

\par\medskip

\newpage

\setcounter{page}{1}

\section{Introduction}

For highly transmissible infectious diseases like measles, meningitis, and \mbox{Covid-19}, vaccine hesitancy represents a major threat to public health~\cite{Dube2013,WilderSmith2019,Siciliani2020,Bozzola2020}, preventing countries from reaching herd immunity. The reasons for people deciding not to get vaccinated (or not to vaccinate their children) range from underestimating the risk of contracting the disease and overemphasizing the vaccine's potential side-effects to general distrust in medical professionals and public health officials \cite{MacDonald2015,Jarrett2015,Storr2018,Shen2019,Quinn2019}. In order to develop effective policy tools to tackle vaccination hesitancy, it is therefore of prime importance for public health authorities to understand the dynamics of an anti-vaccine opinion in the population. In this paper, we introduce a novel mathematical model of opinion dynamics with spatial reinforcement, which allows for the emergence of echo chambers, i.e. opinion bubbles in which information that is incompatible with one's entrenched worldview, is likely disregarded. 

The mathematical modeling of vaccine hesitancy has a long tradition. Similar to the incidence of an infection, vaccination decisions are based on social behavior (see, e.g., the review articles~\cite{Funk2010,Yaqub2014}). However, there is a fundamental difference: whereas getting infected is a function of exposure and (bad) luck, refusing or accepting vaccination is -- provided sufficient supply of vaccines -- a deliberate decision. Since the seminal work of Fine and Clarkson~\cite{Fine1986}, a game-theoretical approach prevails. The individual aims to minimize their personal risk whereas public health authorities wish to improve the overall health status of the population (e.g., minimizing the disease burden~\cite{McDonald2012}). When these two objectives stand in conflict (e.g. if side-effects of a vaccine are known or presumed), individuals may refuse vaccinations which -- from a public health perspective -- would be undesirable~\cite{Fine1986,Mueller1997,Bauch2003,Galvani2007,Fu2010}.

A more recent strand of literature on vaccine hesitancy bypasses utilitarian considerations and focuses directly on opinion dynamics  \cite{Salathe2008,AlvarezZuzek2017,Pires2018}. Rather than modeling individual decisions as a result of utility maximization, the individual is seen as being influenced by the social environment; e.g. the behavior of one's fellow citizens, news coverage, and public health recommendations. This framework has been taken up by the social learning theory for vaccine hesitancy~\cite{Basu2008,Bauch2012,moebius2007}. These papers either focus on data analysis, or use generic models to address social learning. Most papers addressing the link between opinion dynamics and vaccination hesitancy~\cite{Eames2009,AlvarezZuzek2017,VelasquezRojas2017,Bhattacharyya2019} do not validate their models by data analysis. A notable exception is \cite{Salathe2008}, which shows that spatial clusters of local outbreaks can be explained by clusters of low vaccination coverage, presumably caused by local opinion dynamics. However, extant models lack an underlying mechanism for opinion dynamics and its change over time.

More generally, the presence and dynamics of vaccine hesitancy in a population is a specific case of modeling opinion dynamics over time. The basis of all opinion models, the voter model~\cite{Liggett1985}, has the disadvantage that in a finite, well-mixed population, only one opinion remains in the long run, which obviously is not in line with the empirical reality. A remedy is the introduction of zealots or stubborn individuals which decreases, while not fully preventing, the occurrence of opinion loss (~\cite{Mobilia2013,Fernandez-Gracia2014,Braha2017, Galam2007,Galam2015} and the review article~\cite{Dong2018}). Yet, the zealot model does not incorporate mechanisms that lead to the formation of echo chambers. Instead, nonlinear opinion models may account for such effects (see, e.g.\  Sznajd-models~\cite{Sznajd-Weron2005,Sznajd-Weron2000}, or the Ising model~\cite{Nicolao2019}).

While extant nonlinear models are rather phenomenological and not behaviorally validated by empirical studies, the spatial reinforcement model introduced here aims to implement behavioral mechanisms which have been extensively documented in psychology, economics, and political science. In our model, zealots represent people's exposure to stubborn individuals, i.e. characterized as exhibiting partisan (pro-vaccine or anti-vaccine) information. Controlled behavioral experiments show that distorted information critically affects people's decision making such as the willingness to cooperate \cite{engel2021managing}. During the current \mbox{Covid-19} crisis, higher exposure to Fox News has been found to reduce people's propensity to follow stay-at-home orders \cite{simonov2020persuasive}.

In addition, our model accounts for the fact that individuals are disproportionately exposed to ideas similar to their own. In other words, their ideas are being reinforced. Behavioral research has documented two main sources of reinforcement. On the one hand, there are technical filters of information. Both traditional media outlets and social networks have incentives to align their stories to the (presumed) opinions of their readers and users~\cite{Sunstein2001,Pariser2011,enke2020you}. However, even in social networks individuals are usually exposed to a large range of information ~\cite{Borgesius2016,Flaxman2016}. In addition to technical filters, information is being filtered -- often subconsciously -- by the individuals themselves \cite{Festinger1957,Barlett1932,Taylor1981}. People are more likely to dismiss information that is in conflict with one's entrenched worldview \cite{kahan2012ideology,fryer2019updating}, as well as information from sources that are perceived as outgroup \cite{Tajfl1986,Cohen2003}. In particular, political partisanship has been found to have a sizable effect on people's acceptance of government vaccination recommendations. Democrats were much more likely than Republicans to believe in the safety of the swine flu (H1N1) vaccine, introduced in 2009 by the Obama administration. In contrast, Republicans were much more likely than Democrats to believe in the safety of the smallpox vaccine, introduced in 2003 by the Bush administration \cite{krupenkin2021does}. In the current \mbox{Covid-19} crisis, Democrats are more likely to believe in the perils of contracting the virus, and in the effectiveness of social distancing \cite{allcott2020polarization}.

In the methods section, we summarize for non-mathematical readers the main principle of our spatial reinforcement model. We then provide a more detailed description of the model and of the vaccination coverage data used. In the results section, we first analyze the model behavior analytically by scaling to a deterministic limit and discuss the bifurcation structure in a relatively simple two-patch system. Subsequently, we scale the model to a weak effects limit, which yields a Fokker-Planck equation, and obtain the invariant measure. In the second part of the results section, we use this invariant measure to analyze vaccination data for measles and meningococci in Germany. We show that the model is appropriate, and use the estimations as a basis to predict the effectiveness of various policies aimed at increasing vaccination uptake.

\section{Methods}
\paragraph{Mathematical model in a nutshell.}
To study the opinion dynamics of vaccine hesitancy, we develop a spatial, i.e. graph-based, reinforcement model which allows for the emergence of echo chambers. These are environments in which individuals' (pro-vaccine or anti-vaccine) opinions get reinforced through biased interaction with peers or exposure to sources with similar tendencies and attitudes as themselves \cite{Cinelli2021}. Our spatial reinforcement model builds on the classic voter model~\cite{Liggett1985}. Each individual is assumed to be born with a certain opinion (pro-vaccine or anti-vaccine). In order to change their opinion, individuals need to be exposed to the opposite opinion. In this case, an individual has a certain probability to change their current opinion.

We introduce three important extensions to the voter model. First, we allow for the presence of so-called zealots (~\cite{Mobilia2013,Fernandez-Gracia2014,Braha2017, Galam2007,Galam2015}, reviewed in~\cite{Dong2018}). Zealots are agents who never change their mind, and as such are permanent senders of a particular partisan position. They represent the presence of (and exposure to) partisan information sources that persistently advocate in favor or against vaccination (\textit{e.g.} stubborn individuals, national health authorities, newspapers, internet platforms, etc.). Second, we introduce a spatial structure. In the basic form of the voter model used here, interaction happens only within a geographically determined area (also called a patch). In our spatial model, exposure to other opinions is also possible between individuals who live in neighboring patches (in addition to one's own patch). Third, we allow for reinforcement of current opinions. Behavioral research documents that humans are more likely to engage with people who have similar behavior or belief as themselves, and that they tend to discredit opinions that are not aligned with their own \cite{iyengar2019origins}. In our model, reinforcement enables the emergence of echo chambers. 

We study the dynamics of opinion, i.e. the frequency of a certain opinion -- pro-vaccine or anti-vaccine -- per patch and across patches. As our model is stochastic by nature, we first analyze a deterministic limit which explicit the long term behavior of the model, that is whether one or both opinion can be maintained and what is the expected frequency of each opinion in each patch. Second, we use a weak-effects limit to obtain the expected distribution of opinion frequencies within and across patches. This so-called invariant measure will later be used for data analysis.

\paragraph{Model description.} 
We define $\Gamma$ as an undirected graph. We write $k\in\Gamma$ to indicate that $k$ is a node of the graph, with each node representing a geographic unit, \textit{e.g.}\ a patch or in our real data a district. For two nodes $k,k'\in\Gamma$ we define the relation $k\sim k'$ if and only if there is an edge between $k$ and $k'$. Let $d_k$ denote the degree of a given node $k\in\Gamma$, that is the number of neighboring nodes. The nodes (patches/districts) themselves are assumed to have an identical social structure, and an identical population size. Let $N$ denote the total population size in one district, $N_i$ the number of zealots with opinion $i\in\{1,2\}$, and $n_i=N_i/N$ the share of opinion $i$ zealots in a given district. In our vaccination data, these represent stubborn pro-vaccine and anti-vaccine opinions, respectively. Furthermore, let $X_t^{(k)}$ be the number of (non-stubborn) opinion~1 supporters (pro-vaccine) in node $k\in\Gamma$ at time $t$, while $N-X_t^{(k)}$ is the number of (non-stubborn) opinion~2 (anti vaccine) supporters.

Individuals change their opinion by being exposed to other opinions. In the basic zealot model, a given individual changes opinion at rate $\mu$. In that process, the individual simply interacts with a randomly chosen other person (including the zealots) from that district, and adopts that person's opinion. In order to take interaction (and thus potential opinion spillovers) between neighboring patches/districts into account, we define the convex combination of the number of opinion~1 supporters in a given district $k$ and the average number of opinion~1 supporters in the neighboring districts:
\begin{eqnarray}
\hat X_t^{(k)} = (1-\tau)X_t^{(k)}+\tau \check X^{(k)},\qquad \check X^{(k)} := \frac 1 {d_k} \sum_{k' \sim k} X_t^{(k')}.
\end{eqnarray}
The parameter $\tau$ represents the strength of connectedness between neighboring districts. A person meets with probability $1-\tau$ an individual from their own district, and with probability $\tau$ an individual of a randomly chosen neighboring district. A person thus interacts with an opinion 1 supporter (and thus adopts that opinion) with probability
$$ \frac{\hat X_t^{(k)}+N_1}
{N_1+N_2+N}.$$

\noindent We define the incremental increase or decrease, respectively, of the number of opinion 1 supporters as the spatial zealot model:
 \begin{eqnarray}
 	X_t^{(k)}\rightarrow X_t^{(k)} + 1 &\mbox{ at rate } &  \mu (N-X_t^{(k)})\,\frac{\hat X_t^{(k)}+N_1}
 	{N_1+N_2+N},\label{zealotSpaceParticle1}\\
 	X_t^{(k)}\rightarrow X_t^{(k)} - 1 &\mbox{ at rate } &  \mu X_t^{(k)}\,\frac{N-\hat X_t^{(k)}+N_2}
 	{N_1+N_2+N}.\label{zealotSpaceParticle2} 
 \end{eqnarray} 
 \newline
A person in an echo chamber is less likely to flip to the opposite opinion, be it because it is less likely to meet people of the opposite opinion, or because that person discards the alternative opinion as nonsense. We subsume both reasons as a decrease in the effectiveness of interaction. Let $\vartheta_1$ denote the probability for a opinion-2 person to interact effectively with an opinion-1 supporter. An opinion-2 supporter interacts effectively with all opinion-2 supporters ($1-\hat X_t^{(k)}+N_2$), and with the fraction $\vartheta_1$ of opinion-1 supporters ($\vartheta_1(\hat X_t^{(k)}+N_1)$). With $\vartheta_1$ capturing the strength of reinforcement of opinion 2 supporters, the probability of an opinion-2 supporter becoming an opinion-1 supporter reads
$$
\frac{\vartheta_1 (\hat X_t^{(k)}+N_1)}
{\vartheta_1(\hat X_t^{(k)}+N_1)+(N-\hat X_t^{(k)}+N_2)}.
$$
Similarly, we introduce $\vartheta_2$ as the probability for a opinion-1 supporter to interact effectively with a opinion-2 person. 
With this idea, we can define the increase and decrease, respectively, of the number of opinion 1 supporters  as the spatial reinforcement model:
\begin{eqnarray}
	X_t^{(k)}\rightarrow X_t^{(k)} + 1 &\mbox{ at rate } &  \mu (N-X_t^{(k)})\,\frac{\vartheta_1 (\hat X_t^{(k)}+N_1)}
	{\vartheta_1(\hat X_t^{(k)}+N_1)+(N-\hat X_t^{(k)}+N_2)},\label{reinfSpaceParticle1}\\
	X_t^{(k)}\rightarrow X_t^{(k)} - 1 &\mbox{ at rate } &  \mu X_t^{(k)}\,\frac{\vartheta_2 (N-\hat X_t^{(k)}+N_2)}
	{(\hat X_t^{(k)}+N_1)+\vartheta_2 (N-\hat X_t^{(k)}+N_2)}.\label{reinfSpaceParticle2}
\end{eqnarray} 
In the analytical part of the results we scale the model both to a deterministic limit and to a weak-effects limit, in order to obtain bifurcations, phase transitions, and the invariant measure.

\paragraph{Vaccination data.} 
We apply the spatial reinforcement model to data on vaccination coverage for measles in Germany for child cohorts born in 2008-2012, by district and birth year. For those birth years, a district's measles vaccination quota was measured using a consistent method across Germany, and is publicly available through the Robert-Koch-Institute (RKI) (\cite{Goffrier2016,Schulz2013} and supplementary material). Measles is a highly infectious childhood disease \cite{Guerra2017}, with rare but serious complications, such as measles pneumonia. The standard public health recommendation in Germany is a first measles vaccination for children at an age of 11-14 months, and a second shot at 15-23 months~\cite{rkiRatgeberMasern}. We focus our analysis on the first vaccination shot because it best captures deliberate vaccination denial (rather than negligence/forgetfulness). Since public health recommendations are clear and salient, and access to the vaccine is convenient and free of charge, we can consider parents' decision to not have their child vaccinated as a direct reflection of their opinions towards vaccination (pro- or anti-vaccine). In the SI Appendix, we present a similar analysis using data on meningococci vaccinations.

\paragraph{Model fitting and statistical analysis.} 
In order to estimate the parameters of our spatial reinforcement model using the measles or meningococci vaccination data across all 413 districts in Germany, we use a statistical method based on likelihood computation. Data analysis is performed in two steps. First, we estimate the parameters of the decoupled model (assuming $\tau=0$) and compare its empirical fit with and without reinforcement. We compute the exact likelihood ${\mathcal {L}} = P(x^{(k)}=y^{(k)})$, with $x^{(k)}$ being the fraction of vaccinated individuals in district $k\in\Gamma$. Second, we estimate the parameters of the connected model ($\tau\geq 0$). In that case, we observe the data $y^{(k)}$ in grid node $k$, but obtaining the exact likelihood comes at prohibitively high computational costs. Following~\cite{Frank1986,Strauss1990,Anderson1999}, we approximate the likelihood by the product of the marginal probabilities for single nodes, conditioned on the state of all other nodes. This conditioned-likelihood (pseudo-likelihood) allows us to compute the normalizing constant $C$ by a one-dimensional integration ($C$ is a key parameter defined in the proposition and theorem in the results section). Note that this integration has to be performed for each node, but it is still faster and more practical to compute $|\Gamma|$ one-dimensional integrals rather than one $|\Gamma|$-dimensional integral. Parameter estimations based on pseudo-likelihood and exact likelihood typically yield comparable results~\cite{Strauss1990}. The general pseudo-likelihood formula can be written as follows:
$$\widehat{\mathcal{L}} = \prod_{k\in\Gamma} P(x^{(k)}=y^{(k)}\,|\,x^{(k')}=y^{(k')},\,\,k'\not = k).$$

Finally, in order to asses the sensitivity of the vaccination coverage $v$ to changes in the model parameters $p$, i.e. zealot strength ($N_1$, $N_2$) and reinforcement strength ($\vartheta_1$, $\vartheta_2$), we compute the respective elasticities (for a similar approach, see \cite{Zi2011}): $$ e_{v,p} = \frac{p}{v}\,\frac{\partial v}{\partial p}.$$
The elasticity $e_{v,p}$ is dimensionless and approximates the percentage change in vaccination coverage $v$ in response to a percentage change in the respective parameter $p$. \newline

\section{Results}
\subsection{Analytical results}
\paragraph{Deterministic limit.} To understand the general behavior of the model, we first scale the model to a deterministic continuum limit, using the assumption that the population size of the nodes (districts) is large. In this case, the stochastic property of the model becomes negligible and we obtain the deterministic limit, whose resulting equation can be readily interpreted. 

Before analyzing the more complex two-patch systems, let us briefly consider the simplest possible case, namely a single patch with symmetric parameters (same number of zealots and same strength of reinforcement for both opinions). We show that the, if reinforcement becomes strong enough, patch undergoes a so-called pitchfork bifurcation, meaning that three outcomes can occur: opinion 1 dominates, opinion 2 dominates, and both opinions coexist at intermediate frequencies in the patch. More precisely, with low levels of reinforcement, we find only one locally asymptotically stable state, in which one of the two opinions vanishes. In contrast, when reinforcement is high, we find three branches of stationary states, of which two are locally asymptotically stable (the single opinion states), and the middle one (coexistence of both opinions) is unstable.

\begin{figure}[H]
	\begin{center}
	\includegraphics[width=\textwidth]{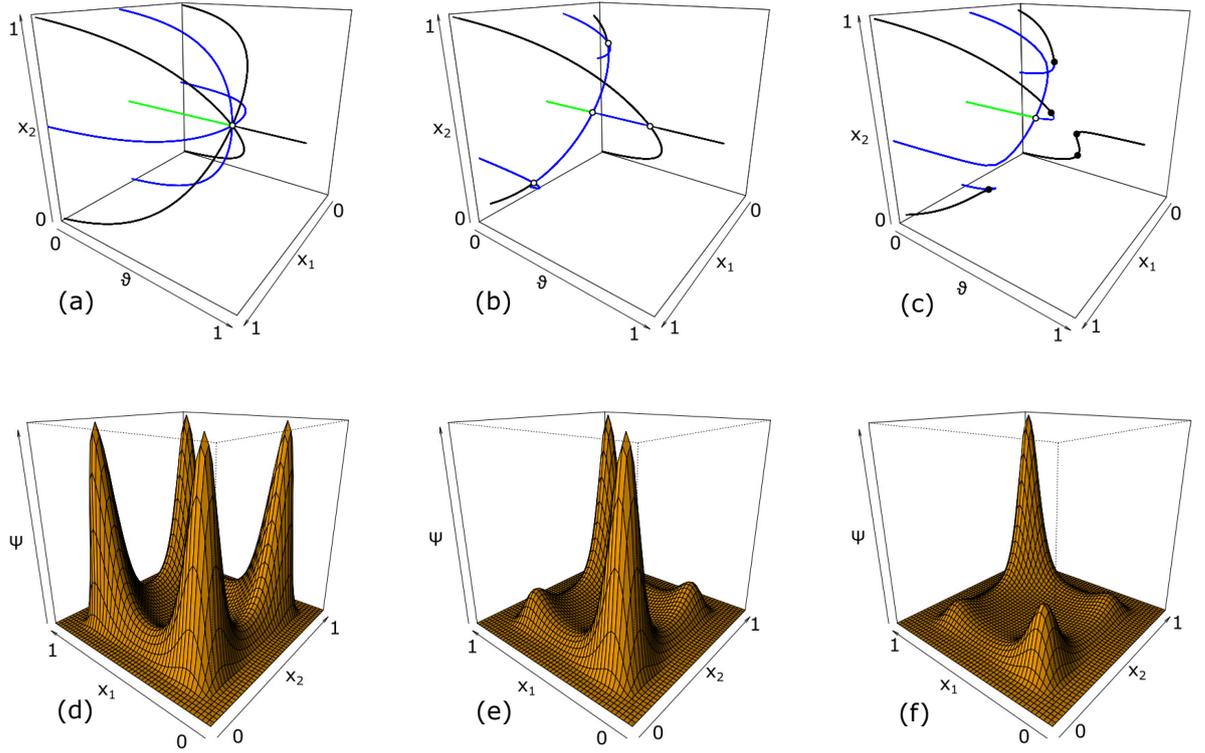}
	\end{center}
	\caption{\underline{Upper row:}  Stationary states for the deterministic two-patch model in dependency of the reinforcement parameter $\vartheta$, with 	$\vartheta_1=\vartheta_2 = \vartheta$. (a) Symmetric and decoupled $n_1=n_2=0.1$, $\tau=0$. (b) Symmetric and connected, $n_1=n_2=0.1$, $\tau=0.1$. (c) Non-symmetric and connected, $n_1=0.1$, $n_2=0.105$, $\tau=0.1$. Black: locally asymptotically stable states are in black; blue/green: unstable states states (1- and 2-dimensional unstable manifold), closed circles: saddle-node bifurcations, open circles (degenerated) pitchfork circles.\\
	\underline{Lower row:} Invariant measure $\psi$ for the stochastic two-patch model, with 	$\theta_1=\theta_2 = 120$. (d) Symmetric and decoupled $N_1=N_2=20$, $\gamma=\tau=0$. (e) Symmetric and connected, $N_1=N_2=20$, $\gamma=20$. (f) Non-symmetric and connected, $N_1=20$, $N_2=20.4$, $\gamma=20$.}\label{invMeasureFigs}
\end{figure}

These results are first extended to two patches without any interaction between them ($\tau=0$) and the outcome is illustrated in Fig. \ref{invMeasureFigs} (a)-(c). The stationary states for the deterministic two-patch model is shown in dependency of the reinforcement parameter $\vartheta$. Locally asymptotically stable (unstable) states are shown in black (blue: one-dimensional, green: two-dimensional unstable manifold), and bifurcation points are depicted as dots. A product structure for the stationary states is found with each combination of single-patch stationary states yielding a valid stationary state for the two-patch situation. As three stationary states occur after the pitchfork bifurcation in a single patch, we find nine branches in the two-patch model after a highly degenerated pitchfork bifurcation. As both single-patch stationary states have to be locally stable to be combined into a locally stable two-patch state, we obtain four branches of stable stationary points: Two of them are convergent (the same opinion prevails in both patches) and the other two are divergent (opposite opinions prevail in the two patches).

We can now extend the deterministic limit analysis to the two-patch model with interaction ($\tau>0$). The comparison between panels (a) and (b) of Fig.~\ref{invMeasureFigs} illustrates that as interaction between patches increases ($\tau=0.1$), the degeneracy decreases (still under an identical set of parameter values for both opinions). The convergent stationary state $x_1=x_2=1/2$ undergoes two subsequent non-degenerate pitchfork bifurcations. Particularly, the stable part of the divergent stationary solutions becomes smaller. If between-patch interaction increases even further, the divergent solutions eventually vanish as the interaction enforces a convergence of opinions. In other words, when patches are connected, one of the two opinions prevails in both patches. If we break the symmetry of parameter values between opinions (Fig.~\ref{invMeasureFigs} (c)), the pitchfork bifurcations are unfolded into a series of saddle-node bifurcations. In view of the multi-stable situation characterizing the deterministic behavior, we expect -- when accounting for stochasticity -- the invariant measure to show a multi-modal distribution.

\paragraph{Weak-effects limit.} We now present results for the stochastic model under the weak-effects limit. We aim a the analysis of patch data, where each patch has a certain population size. Experience e.g.\ in population genetics~\cite{tavare:book,etheridge} indicates that this limit is better suited for the analysis of such data than the deterministic limit.  For the weak-effects limit, we choose a different scaling of the parameters than before. We assume that the number of zealots $N_i$ are independent of the population size $N$, and that the reinforcement component becomes small for large $N$, $\vartheta_i=1-\theta_i/N$. Note that in the original scaling, a small value of $\vartheta_i$ indicates a strong reinforcement, while from now on (with the new scaling), large values of $\theta_i$ indicate strong reinforcement. 
We furthermore allow the spatial interaction parameter $\tau$ to depend on $1/N$ but also on the frequency of a certain opinion in a given patch, with the new parameter $\gamma$ as proportionality constant,
$$ \tau(x^{(k)}) = \frac 1 N\,\,\gamma\,\, x^{(k)}\,\,(1-x^{(k)}).$$

Interaction \textit{per se} is assumed to be a weak effect, so that $\tau(x^{(k)})$ tends to zero if $N$ tends to infinity. Moreover, if almost all individuals in a given patch have the same opinion ($x^{(k)}\approx 1$ or $x^{(k)}\approx 0$), interaction has almost no effect. In other words, the effect of interaction between patches is maximized if a patch is maximally heterogeneous, i.e. if $x^{(k)}\approx 1/2$. Our assumptions reflect the application of the model. Put simply, we assume that if all my neighbors have the same opinion as myself, I am inclined to follow that opinion too, and do not look further to crosscheck my opinion with people from other districts (interaction parameter $\tau$ is very small). By contrast, if my neighbors give me contradicting pieces of advice, I may be more inclined to ask additional people. Consequently, I am more likely to also communicate with individuals from neighboring patches ($\tau$ is larger). Moreover, our choice of $\tau(.)$ is mathematically convenient as it yields a consistent scaling and an appropriate invariant measure.

Before we investigate the connected model (with interaction between neighboring patches), we first derive the invariant distribution for the decoupled model (without interaction between neighboring patches). While the proof of the next proposition can be found in~\cite{Mueller2020}, we sketch the proof in the SI Appendix for the completeness.

\begin{prop}
	Let $N_i$ denote the number of zealots for group $i$, $N$ the population size, and $\vartheta_i=1-\theta_i/N$ the parameter describing reinforcement. We assume no interaction, $\gamma=0$, s.t.\ all patches become independent and identical. In the limit $N\rightarrow\infty$, the density of the invariant measure for the random variable $x_t=X_t^{(k)}/N$ (for any $k$) is given by 
\begin{eqnarray}
	\varphi(x) =  
	C\,e^{\frac 1 2 (\theta_1+\theta_2)x^2-\theta_1\,x}\,\, x^{N_1-1}\,(1-x)^{N_2-1},\label{reinfDirichDist}
\end{eqnarray}
where $C$ is determined by the condition that the integral over $\varphi(.)$ is $1$.
\end{prop}

Using this result, we can determine the invariant distribution for the connected model, yielding our main theorem below. We are particularly interested in the invariant measure $\psi(.)$ which summarizes the pseudo-equilibrium distribution of the frequencies of opinions in a stochastic model. 

\begin{theorem}
	Let $N_i$ denote the number of zealots for group $i$, $N$ the population size, and $\vartheta_i=1-\theta_i/N$ the parameter describing reinforcement. We also scale the spacial interaction strength $\tau=\frac 1 N\,\,\gamma\,\, x^{(k)}\,\,(1-x^{(k)})$. In the limit $N\rightarrow\infty$, the density of the invariant measure for the random variable $x^{(.)}_t = (x_t^{(1)},\ldots,x^{(|\Gamma|)})=(X_t^{(1)}/N,\ldots,X_t^{(|\Gamma|)}/N)$  is given by 
	\begin{eqnarray}
		\psi(x^{(\cdot)}) =  C\,\prod_{k\in\Gamma}\bigg( \,\,\varphi(x^{(k)})\,\,\,\,
		\exp\bigg\{-\,\frac{\gamma}{4\,d_k}
		\sum_{k'\sim k}(
		x^{(k)}- x^{(k')})^2\bigg\}
		\,\,\bigg),\label{reinfDirichDistSpace}
	\end{eqnarray}
	where $\varphi(.)$ is the homogeneous-population distribution defined in eqn.~(\ref{reinfDirichDist}), and $C$ is determined by the condition that the integral over $\psi(.)$ is $1$.
\end{theorem}

The distribution of opinion frequencies has a multiplicative structure, where the first term is related to the local dynamics within a given patch $k$ ($\varphi(x^{(k)})$), while the second term accounts for the communication between neighboring patches. For the two-patch system, the shape of the invariant distribution can be visualized (Fig.~\ref{invMeasureFigs}). The local maxima are the stochastic analogue of the locally asymptotically stable stationary points in the deterministic case. In fact, Fig.~\ref{invMeasureFigs} (d)-(f) correspond to Fig.~\ref{invMeasureFigs} (a)-(c) for appropriate choices of $\vartheta$. We represent in subfigure (d) the case without patch interaction for which four local maxima appear, mimicking the deterministic results of four stable stationary points. If interaction strength increases ($\gamma>0$ defining $\tau$), the non-symmetric maxima decrease, while the symmetric maxima (on the line $x_1=x_2$) are still present (subfigure (e)). If we choose non-symmetric parameters ($N_1\not = N_2$ in subfigure (f)), we still find the local maxima on the diagonal, but now one maximum dominates the distribution. To conclude, the different parameters (strength of reinforcement, strength of zealots, interaction between patches) strongly influence the shape of opinion frequency distributions. All parameters being identifiable, we set to estimate the parameters of our model based on real-world data. 

\subsection{Vaccination data analysis}
\begin{figure}[H]
	\begin{center}
\includegraphics[width=\textwidth]{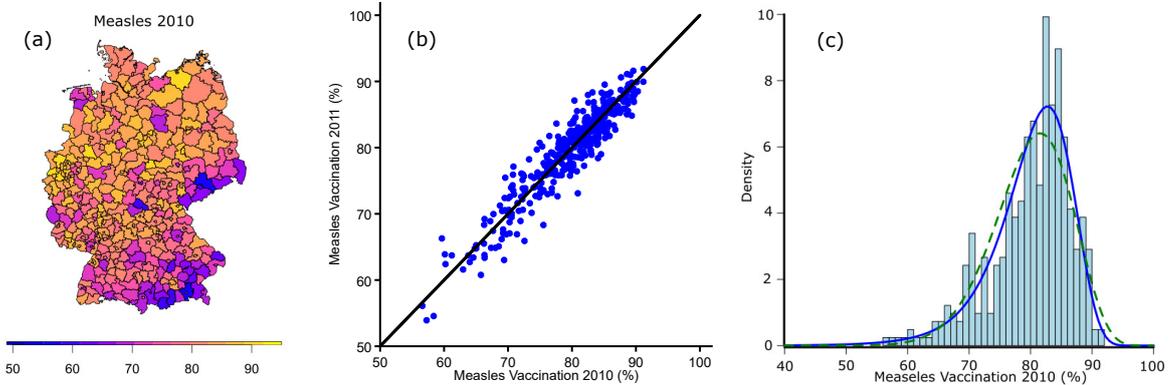}
	\end{center}
	\caption{\underline{Panel (a):} Measles vaccination coverage (in percent) for children born in 2010 across 413 districts in Germany (GIS-data: \textcopyright GeoBasis-DE / BKG (2021)). \underline{Panel (b):} Vaccination coverage for birth years 2010 and 2011, by district; 45 degree line for comparison. \underline{Panel (c):}  Histogram of vaccination coverage for measles per district in 2010. The solid blue (dashed green) line: fit of the decoupled model with (without) reinforcement. Bin size is $0.01$.} \label{measlesGIS2012}
\end{figure}

The RKI defines a district's vaccination quota as the number of children born in year X getting their first measles shot within their first 24 months, divided over all children born in year X. The left panel of Fig.~\ref{measlesGIS2012} illustrates the variation of vaccination quotas across Germany. Visually, the data suggests the existence of spatial correlations, \textit{e.g.} a large cluster with a low vaccination coverage in southern Germany. Moreover, as shown in the middle panel of Fig.~\ref{measlesGIS2012}, districts' vaccination quotas are strikingly consistent across birth years. In our dataset, the smallest year-to-year correlation is $\rho=.87$ (Spearman correlation coefficient). This resonates with our assumption that people's local environment is central for their opinion formation and thus for their decision to support or reject vaccination. 

We assess the extent to which reinforcement explains the distribution of vaccine opinions across districts in Germany by estimating the reinforcement parameter $\theta_i$. Fig.~\ref{measlesGIS2012} (right panel) illustrates the superior fit of the decoupled model with reinforcement rather than without reinforcement. The model without reinforcement is rejected (Kolmogorow-Smirnow-test, significance levels $p<0.05$ for all birth years separately). In contrast, the model with reinforcement is appropriate for the data (Kolmogorow-Smirnow-test, significance levels $p>0.05$ for all birth years separately), and the reinforcement component is indispensable for explaining the data (the likelihood-ratio test rejects the model without reinforcement, $p<0.000005$, see SI Appendix for details).

\begin{figure}[H]
	\begin{center}
		\includegraphics[width=\textwidth]{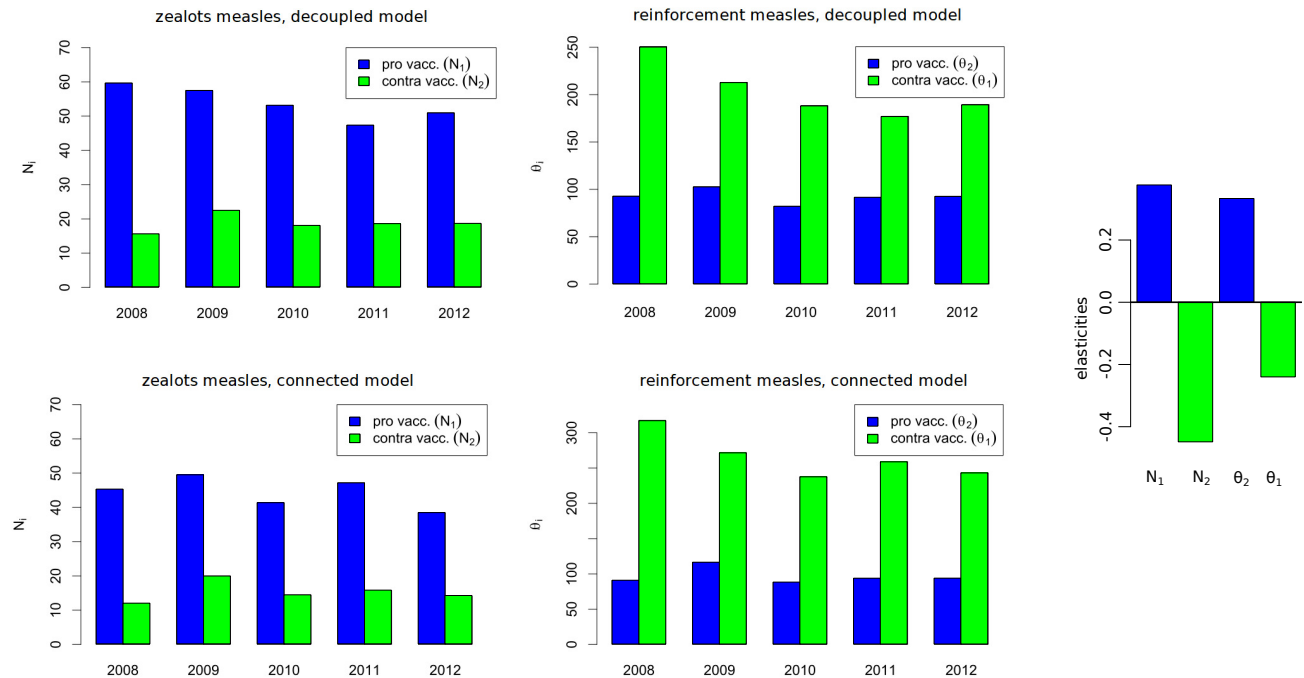}
	\end{center}
	\caption{Estimation parameters for the decoupled model (upper panels) and the connected model (lower panels).  Right panel: Elasticities for the decoupled model (parameters of measles 2010). Blue: pro-vaccination parameters, green: anti-vaccination parameters.}\label{HomoResultFig}
\end{figure}

The upper panels of Fig.~\ref{HomoResultFig} display the estimated model parameters for the decoupled model, for each birth year separately. We find the zealot parameter to be larger for the pro-vaccination than for the anti-vaccination opinion. In contrast, the reinforcement parameter is larger for the anti-vaccination than for the pro-vaccination opinion. This pattern is consistent across all birth years. Additionally allowing for interaction between neighboring districts is meaningful (the estimated $\gamma$ is consistently positive) but does not affect the general patterns found above. In fact, the upper panels (decoupled model) and the lower panels of Fig.~\ref{HomoResultFig} (connected model) are striking similar. The zealot component is consistently larger for the pro-vaccination opinion whereas the reinforcement component is larger for the anti-vaccination opinion. This suggests that individuals are more often exposed to information that promotes vaccination, but vaccination deniers are more likely to filter out this pro-information and focus on anti-information. In the SI Appendix, we show that the relative sizes of all four parameters are robust to applying the model to meningococci vaccinations rather than measles.

As shown in Fig.~\ref{HomoResultFig}, right panel, we find as intuitively expected, that increasing any of the two pro-vaccination parameters ($N_1$, $\theta_2$) leads to an increase of the vaccination coverage, while increasing the anti-vaccination parameters ($N_2$, $\theta_1$) entails a decrease. More interestingly, we find that decreasing the zealot strength of vaccination-deniers $N_2$ has the strongest effect on vaccination uptake. In contrast, influencing the reinforcement parameter of vaccination deniers has the smallest impact.

\section{Discussion}

We have developed and analyzed a novel spatial opinion model that incorporates reinforcement, thus allowing for the occurrence of echo chambers. Applying our model to fit and draw parameter inference from vaccination data for measles and meningococci, we find our novel reinforcement model to be highly superior to a model without reinforcement. Estimated parameters are robust across both diseases and all birth years in our sample, and to allowing for spatial correlations between neighboring districts. 

The model introduced here is the first model capable of structuring the factors influencing vaccination hesitancy into (i) exposure to -- potentially distorted -- partisan information and (ii) people's tendency to consume information that is aligned with their worldview; and to allow these components to be measured with spatial data. We find the zealot component to be much larger for the pro-vaccine opinion than for the anti-vaccine opinion. This resonates with the idea that in Germany the pro-vaccine opinion has much more exposure than the anti-vaccine opinion. In contrast, the reinforcement component is much larger for the anti-vaccine opinion. This is consistent with empirical studies that associate vaccine hesitancy with people's receptiveness to populism and conspiracy theories~\cite{Kennedy2019,Mueller2020,Stecula2021}.

By computing elasticities, our model can be used to predict the effectiveness of different measures aiming at increasing vaccination uptake. We find that the most sensitive parameter influencing vaccination uptake is exposure to the anti-vaccine opinion. According to our model, measures aiming at reducing the salience of partisan anti-vaccine information sources would have the largest effect on enhancing vaccination uptake. In contrast, we find that measures aiming at reducing the reinforcement of vaccination deniers have the smallest impact. This resonates with the idea that dialogue-based approaches which are carefully targeted to a specific social group are among the most effective measures against vaccination hesitancy~\cite{Jarrett2015}. Studying vaccine hesitancy for \mbox{Covid-19}, Kl\"uver et al.~\cite{Kluever2021} find positive information, appropriate communication, and trust to be central. If the behavioral mechanisms behind people's reluctance to receive a \mbox{Covid-19} vaccination are similar to their opposition to vaccines against measles and meningococci, our study suggests that targeting the partisan spreaders of anti-vaccine information could be very effective. An important difference between \mbox{Covid-19} vaccinations on the one hand, and measles and meningococci vaccinations on the other hand, is the incidence of the disease. Both measles and meningitis have been extremely rare in Germany for many years. Consequently, our model neglects the influence of the disease incidence on the vaccination rate, as well as the effect of the vaccination coverage on the infection dynamics~\cite{Bauch2012}. In future research, our model could be generalized to investigate vaccination opinion dynamics for prevalent diseases.

Beyond vaccination hesitancy, our model is applicable to a large range of problems of opinion polarization; from discrepancies about the extent and causes of global warming, to perceptions of inequality and the role of government \cite{egan2017climate,alesina2020polarization,kurschilgen2021moral}. In psychology, economics, and political science, a prominent explanation for polarization has been the idea of motivated reasoning \cite{kahan2012ideology,benabou2016mindful,zimmermann2020dynamics}: When people are emotionally invested in a certain state of the world being true -- e.g. because it favors them economically, or because their partisan alignment is a critical part of their identity \cite{iyengar2019origins} -- it limits their ability to process information in an unbiased manner. While our reinforcement model captures the behavioral biases of the receivers of information, it neglects the motives of the senders of information \cite{kurschilgen2019communication,gneezy2020bribing}. In a world that relies increasingly on highly specialized expert knowledge, while at the same time citizens’ trust in experts is being constantly undermined by populist leaders and media outlets, future work should expand our model to capture the interplay between expert senders and non-expert receivers of information.

Finally, we comment on a more technical aspect of our model. We introduced the model as a branching process at the level of single individuals. Looking at a deterministic limit, we found a sequence of pitchfork bifurcations. Given appropriate parameters, these bifurcations collapsed into a highly degenerate pitchfork bifurcation at a single point. This bifurcation structure resembles that of the Ising/Curie-Weiss model, which has also been applied to opinion dynamics~\cite{Nicolao2019}. Yet, the central mechanism of our model is fundamentally different. While the Ising/Curie-Weiss model implements a majority rule~\cite{Krapivsky2017}, our spatial reinforcement model is -- just as the basic voter model -- based on pairwise interactions. We show that a different scaling of the parameters yields the weak-effects limit. This limit is a stochastic process that is independent of the population size. It is thus suited to analyze data of populations of an appropriate size. Nevertheless, if the population within a patch is too small, the branching process is better suited but stochastic noise can be very strong decreasing the power of parameter inference. If the population is too large, the noise averages out, and we approach a deterministic model. Future work should investigate for which geographic/administrative units our model can be best applied in order to be applicable to different countries and different opinion frequencies.

\bibliographystyle{abbrv}

\bibliography{hesitancy,socialSciencesLit}

\newpage $\quad$\newpage 
\begin{appendix} 
	
\section{Supplementary Material}

\subsection{Analytical results: Deterministic limit}

We present here the results of the deterministic limit.
\begin{theorem}
	Let $x_k(t)=X^{(k)}(t)/N$, $n_i=N_i/N$. Then, the deterministic limit of the model is given by the ODE
	\begin{eqnarray}
	\dot x_k &=& 
	\mu\bigg(
	-\frac{\vartheta_2\,x_k\,((1-\hat x_k)+n_2)}
	{ (\hat x_k+n_1) + \vartheta_2((1-\hat x_k)+n_2)}
	+\frac{\vartheta_1 \,(1-x_k)\,(\hat x_k+n_1)}{\vartheta_1 (\hat x_k+n_1) + ((1-\hat x_k)+n_2)}\bigg)\\
	\hat x_k &=& (1-\tau) x_k + \frac{\tau}{d_k}\sum_{k'\sim k} x_{k'}.
	\end{eqnarray}
\end{theorem}
{\bf Proof: }
Let $X_t=(X_t^{(k)})_{k\in\Gamma}$, and $x(t)=(X_t^{(k)}/N)_{k\in\Gamma}$. The rates to increase/decrease the state $X^{(k)}_t$  can 
be written as
$f_{+,k}(X_t/N)$ resp.\ $f_{-,k}(X_t/N)$, where 
$$f_{+,k}(x) = \mu (1-x_k)\,\frac{\vartheta_1 (\hat x_k+n_1)}{\vartheta_1(\hat x_k+n_1)+(1-\hat x_k+n_2)},\qquad 
f_{-,k}(x) = \mu x_k\,\frac{\vartheta_2 (1-\hat x_k+n_2)}{(\hat x_k+n_1)+\vartheta_2 (1-\hat x_k+n_2)}.$$
$\hat x_k$ is defined in the statement of the theorem. 
Therewith, the Kramers-Moyal expansion yields the limiting Fokker-Planck equation for the probability density $\psi(x,t)$ 
$$ \partial_t \psi(x,t) 
= -\sum_{k\in\Gamma}\partial_{x^{(k)}}( (f_{+,k}(x)-f_{-,k}(x))\,\psi(x,t)) 
+\frac 1 {2N}\sum_{k\in\Gamma}\partial_{x^{(k)}}^2 ( (f_{+,k}(x)+f_{-,k}(x))\,u(x,t)) 
$$
The deterministic ODE is determined by the drift term, such that $\dot x_k = f_{+,k}(x)-f_{-,k}(x)$. 
This yields the desired result.
\qed \par

\subsection{Analytical results: Proof of Proposition}
Proof: We  start  with the Fokker-Planck equation obtained by the Kramers-Moyal expansion, where we use the scaling $n_i=N_i/N$, and $\vartheta_i$ constant in $N$. Only afterwards, we proceed to the desired scaling.\\

As seen above, the rates to increase/decrease the state can 
be written as
$f_+(X_t/N)$ resp.\ $f_-(X_t/N)$, where 
$$f_+(x) = \mu (1-x)\,\frac{\vartheta_1 (x+n_1)}{\vartheta_1(x+n_1)+(1-x+n_2)},\qquad 
f_-(x) = \mu x\,\frac{\vartheta_2 (1-x+n_2)}{(x+n_1)+\vartheta_2 (1-x+n_2)}.$$
As we are in a one-patch-model (no spatial structure), we do not have an index $k$, and also no variable $\hat x$ that is responsible for the communication between patches. Therewith, the limiting Fokker-Planck equation reads
$$ \partial_t u(x,t) 
= -\partial_x( (f_+(x)-f_-(x))\,u(x,t)) 
+\frac 1 {2N}\partial_x^2 ( (f_+(x)+f_-(x))\,u(x,t)) 
$$
Now we rewrite drift and noise term with the new scaling $n_i=N_i/N$, 
$\vartheta_i=1-\theta_i/N$, where we neglect terms 
of order ${\cal O}(N^{-2})$. We find (using the computer-algebra tool maxima~\cite{maxima}) that ($h:=1/N$)
\begin{eqnarray*}
	&&f_+(x)-f_-(x)\\ 
	&=& 
	\mu (1-x)\,\frac{(1-h\,\theta_1) (x+h\,N_1)}{(1-h\,\theta_1)(x+h\,N_1)+(1-x+h\,N_2)}
	-
	\mu x\,\frac{(1-h\,\theta_2) (1-x+h\,N_2)}{(x+h\,N_1)+(1-h\,\theta_2) (1-x+h\,N_2)} \\
	&=& \mu\bigg(
	[(\theta_1+\theta_2)x-\theta_1]\,x\,(1-x)- (N_1+N_2)\,x+N_1
	\bigg)
	\, h+{\cal O}(h^2),
\end{eqnarray*}
while $h(f_+(x)+f_-(x)) = h\,2\,\mu x(1-x)\, + {\cal O}(h^2)$. If we rescale time, 
$T=\mu\, h\,t$, the Fokker-Planck equation becomes
$$ \partial_T u(x,T) = 
-\,\partial_x\bigg\{\,\,\bigg(
[(\theta_1+\theta_2)x-\theta_1]\,x\,(1-x)- (N_1+N_2)\,x+N_1
\bigg)\,\,u(x,T)\,\,\bigg\}
+ \partial_x^2 \bigg\{x\,(1-x)\, u(x,T)\bigg\}.
$$
For the invariant (and stationary) distribution $\varphi(x)$ the flux of that rescaled Fokker-Planck equation is zero, that is, 
$$-\bigg(
[(\theta_1+\theta_2)x-\theta_1]\,x\,(1-x)- (N_1+N_2)\,x+N_1
\bigg) \varphi(x) + \frac d {dx} \bigg(x(1-x)\,\varphi(x)\bigg) = 0.$$
With $v(x) =x(1-x)u(x)$, we have 
$$
v'(x) = 
\bigg(
[(\theta_1+\theta_2)x-\theta_1]\,+ \frac {N_1}x-\frac{N_2}{1-x}\,x\bigg) v(x)
$$
and hence
$$
v(x) = C\,e^{\frac 1 2 (\theta_1+\theta_2)x^2-\theta_1\,x}\,\, x^{N_1}\,(1-x)^{N_2}
$$
resp.\ 
$$
\varphi(x) = C\,e^{\frac 1 2 (\theta_1+\theta_2)x^2-\theta_1\,x}\,\, x^{N_1-1}\,(1-x)^{N_2-1}
$$
\par\qed\par\medskip

\subsection{Analytical results: Proof of Theorem}
\noindent {\bf Proof: } We again start off with the Fokker-Planck equation, obtained by the Kramers-Moyal expansion, where we use the scaling $n_i=N_i/N$, and $\vartheta_i$ constant in $N$. Only afterwards, we proceed to the desired scaling.
As seen above, the rates to increase/decrease the state in site $k$ can 
be written as
$f_+^{(k)}(X_t^{(\cdot)}/N)$ resp.\ $f_-^{(k)}(X_t^{(\cdot)}/N)$, where 
\begin{eqnarray*}
	f_+^{(k)}(x^{(\cdot)}) &=& \,
	\frac{ [\mu (1-x^{(k)})]\,\,\,[\vartheta_1 ((1-\tau)\,x^{(k)}+\tau \check x^{(k)}+n_1)]}{\vartheta_1((1-\tau)\,x^{(k)}+\tau \check x^{(k)}+n_1)+(1-(1-\tau)\,x^{(k)}-\tau \check x^{(k)}+n_2)},\\
	f_-^{(k)}(x^{(\cdot)}) &=&\,
	\frac{[ \mu  
		x^{(k)}]\,\,[\vartheta_2 (1-(1-\tau)\,x^{(k)}-\tau \check x^{(k)}+n_2)]}{((1-\tau)\,x^{(k)}+\tau \check x^{(k)}+n_1)+\vartheta_2 (1-(1-\tau)\,x^{(k)}-\tau \check x^{(k)}+n_2)}.
\end{eqnarray*}
Here, $\hat x^{(k)}$ is the average of $x^{(\cdot)}$ in the neighborhood of $j$, given by the graph $\Gamma$.

Therewith, the flux $j^{(k)}(x^{(\cdot)})$ for the limiting Fokker-Planck equation is defined by
\begin{eqnarray*}
	j^{(k)}(x^{(\cdot)})
	&=& 
	-\bigg(f_+^{(k)}(x^{(\cdot)})
	-
	f_-^{(k)}(x^{(\cdot)})\bigg)u(x^{(\cdot)})\\
	&&+ \frac 1 {2N}\,\partial_{x^{(k)}}
	\bigg\{\bigg(
	f_+^{(k,\ell)}(x^{(\cdot)})
	+
	f_-^{(k,\ell)}(x^{(\cdot)})
	\bigg)	u(x^{(\cdot)})\bigg\}
\end{eqnarray*}
and the Fokker-Planck equation itself reads
$$ \partial_t u(x^{(\cdot)})
= \sum_{k\in\Gamma}\partial_{x^{(k)}}
j^{(k)}(x^{(\cdot)}).
$$

Now we rewrite drift and noise term with the new scaling $n_i=N_i/N$, 
$\vartheta_i=1-\theta_i/N$, $\tau=\gamma/N$, where we neglect terms of order ${\cal O}(N^{-2})$. We obtain (again using the computer-algebra program maxima~\cite{maxima}) that ($h:=1/N$)
\begin{eqnarray*}
	&&f_+^{(k)}(x^{(\cdot)})-f_-^{(k)}(x^{(\cdot)})\\ 
	&=& \,\frac{ \mu \, (1-x^{(k)})\,\,\,[\vartheta_1 ((1-\tau)\,x^{(k)}+\tau \check x^{(k)}+n_1)]}{\vartheta_1((1-\tau)\,x^{(k)}+\tau \check x^{(k)}+n_1)+(1-(1-\tau)\,x^{(k)}-\tau \check x^{(k)}+n_2)}\\
	&&
	-
	\frac{ \mu  \,
		x^{(k)}\,\,\,[\vartheta_2 (1-(1-\tau)\,x^{(k)}-\tau \check x^{(k)}+n_2)]}{((1-\tau)\,x^{(k)}+\tau \check x^{(k)}+n_1)+\vartheta_2 (1-(1-\tau)\,x^{(k)}-\tau \check x^{(k)}+n_2)}\\
	&=& \,
	\mu \bigg( 
	(\check x^{(k)}-x^{(k)})\,\gamma\,\,x^{(k)}\,(1-x^{(k)})\\
	&&\qquad\qquad 
	+ x^{(k)}\,(1-x^{(k)})\,(\theta_2\, x^{(k)} - \theta_1 \,(1-x^{(k)}))
	- (N_2+ N_1) x^{(k)} + N_1
	\bigg)
	\, h+{\cal O}(h^2),
\end{eqnarray*}
while $f_+^{(k)}(x^{(\cdot)})+f_-^{(k)}(x^{(\cdot)}) = 2\,\mu\,x^{(k)}\,(1-x^{(k)})+{\cal O}(h)$. 
Hence, in lowest order, $j^{(k)}(x^{(\cdot)})=0$ reads
\begin{eqnarray*}
	&&\partial_{x^{(k)}}
	\bigg\{\bigg(x^{(k)}(1-x^{(k)})\bigg)	u(x^{(\cdot)})\bigg\}\\
	&=& 
	\bigg( 
	x^{(k)}\,(1-x^{(k)})\,(\gamma (\check x^{(k)}-x^{(k)}) + \theta_2\, x^{(k)} - \theta_1 \,(1-x^{(k)}))
	- (N_2+ N_1) x^{(k)} + N_1
	\bigg)	u(x^{(\cdot)}).
\end{eqnarray*}

For $\gamma=0$, this equation collapses to the case without interaction across patches. This observation motivates us to introduce $v(x^{(\cdot)})$ by 
$$ u(x^{(\cdot)})
=v(x^{(\cdot)})\,\,\prod_{k'\in\Gamma}
\varphi(x^{(k')}).
$$
Therewith, we obtain
\begin{eqnarray*}
	\partial_{x^{(k)}}v(x^{(\cdot)})
	&=&
	\gamma\,(\check x^{(k)}-x^{(k)}) \,\,\, v(x^{(\cdot)})
\end{eqnarray*}
with the solution
$$ 
v(x^{(\cdot)}) 
=
C \exp\bigg\{
\gamma \sum_{k\in\Gamma}\,\bigg(\frac 1 {2\,d_k}
\sum_{k'\sim k}x^{(k)}x^{(k')}
- \frac 1 2 \, (x^{(k)})^2
\bigg)\bigg\}.
$$
The factor $1/(2\,d_k)$ is due to symmetry reasons: Each pair of nodes $(k_1,k_2)$ with $k_1\sim k_2$ appears twice in the sum. We rewrite the sum as follows
\begin{eqnarray*}
	&&\sum_{k\in\Gamma}\,\bigg(\frac 1 {2\,d_k}
	\sum_{k'\sim k}x^{(k)}x^{(k')}
	- \frac 1 2 \, (x^{(k)})^2 \bigg)
	=\frac 1 {2\,d_k}
	\sum_{k\in\Gamma}\,\bigg(
	\sum_{k'\sim k}x^{(k)}x^{(k')}
	- d_k \, (x^{(k)})^2\bigg)\\
	&=&-\frac 1 {2 d_k}
	\sum_{k\in\Gamma}\,\bigg(
	\sum_{k'\sim k}\bigg(-x^{(k)}x^{(k')}
	+  (x^{(k)})^2\bigg)\bigg)
	=-\frac 1 {2\,d_k}
	\sum_{k\in\Gamma}\,\frac 1 2\bigg(
	\sum_{k' \sim k}\bigg(
	(x^{(k')})^2-2x^{(k)}x^{(k')}
	+ (x^{(k)})^2\bigg)\bigg)\\
	&=&-\frac 1 {4\,d_k}
	\sum_{k\in\Gamma}\,\bigg(
	\sum_{k'\sim k}\bigg(
	x^{(k)}- x^{(k')}\bigg)^2\bigg).
\end{eqnarray*}

\par\qed\par\medskip

\subsection{Vaccination data analysis}
The data used are made available by the Robert-Koch-Institute, and have been published in~\cite{Goffrier2016,Schulz2013,LamegoGreinerM2016}. 
The web pages fro the measles data are located in the URL's\\
\url{https://www.versorgungsatlas.de/themen/versorgungsprozesse?tab=3&uid=76&cHash=15379e83482f9325cf011f690c059c26} (accessed 2'nd Mai 2021)\\
\url{https://www.versorgungsatlas.de/themen/versorgungsprozesse?tab=3&uid=43&cHash=31605ab96524f1a101ec8e3aac07e388} (accessed 2'nd Mai 2021)\\
and that for meningoccoci
\url{https://www.versorgungsatlas.de/themen/versorgungsprozesse?tab=4&uid=75&cHash=08f43064449be9c3947a6ffd61e0e09a} (accessed 2'nd Mai 2021)\\

The GIS-data used are available at the URL\\
\url{https://gdz.bkg.bund.de/index.php/default/catalog/product/view/id/773/s/nuts-gebiete-1-250-000-stand-01-01-nuts250-01-01/category/8/?___store=default}\par\medskip 

The R-code used to analyze the data is available on request from the authors.

\paragraph{Measles}
We present here all results for the measles vaccination data.

\textbf{Decoupled model} \newline
\adjustbox{width=\textwidth}{
\begin{tabular}{l|llll|lll|l}
	year & $N_1$ & $N_2$ & $\theta_1$ & $\theta_2$ & LL tst & KS(reinf) & KS (zealot) & mean\\
	\hline
	\input{datAnaMeaslesHomoV1}
\end{tabular}
}

\textbf{Connected model} \newline
\begin{tabular}{l|lllll}
year & $N_1$ & $N_2$ & $\theta_1$ & $\theta_2$ &$\gamma$ \\ 
	\hline
	\input{datAnaMeaslesSpatV1}
\end{tabular}

If we exclude Saxonia (as it has for meningococci slightly different rules for vaccination), we obtain\\
\begin{tabular}{l|lllll}
	 year & $N_1$ & $N_2$ & $\vartheta_1$ & $\vartheta_2$ &$\gamma$ \\ 
	\hline
	\input{datAnaMeaslesSpatNoSaxV1}
\end{tabular}

The parameters are slightly different if we exclude Saxonia, but thee is no fundamental difference to the analysis of all data.

\paragraph{Meningocci Infection}
We present here all results of the Meningocci vaccination. \newline
\textbf{Decoupled model} \newline

\adjustbox{width=\textwidth}{
\begin{tabular}{l|llll|lll|l}
 year & $N_1$ & $N_2$ & $\vartheta_1$ & $\vartheta_2$ & LL tst & KS(reinf) & KS (zealot) & mean\\
\hline
\input{datAnaMeningoHomoV1}
\end{tabular}
}

\textbf{Connected model} \newline

As Saxonia has slightly different recommendations with respect to the meningococci vaccination, these data are left out (and are missing in the original data set).

\begin{tabular}{l|lllll}
	 year & $N_1$ & $N_2$ & $\theta_1$ & $\theta_2$ &$\gamma$ \\ 
	\hline
	\input{datAnaMeningoSpatNoSaxV1}
\end{tabular}

That is, the parameters for the spatial model are slightly different, but there is no fundamental difference to the analysis of all data.

\paragraph{Elasticities}
The table with the elasticities show a common pattern for both diseases and all years.
\par\medskip

\begin{tabular}{ll|llll}
name & year & Elast($N_1$)  & Elast($N_2$)  & Elast($\vartheta_1$)  & Elast($\vartheta_2$)\\
 	\hline
	\input{elast}

\end{tabular}
\newpage

\section{SI Figures}

\begin{figure}[h!]
	\begin{center}
		\includegraphics[width=7cm]{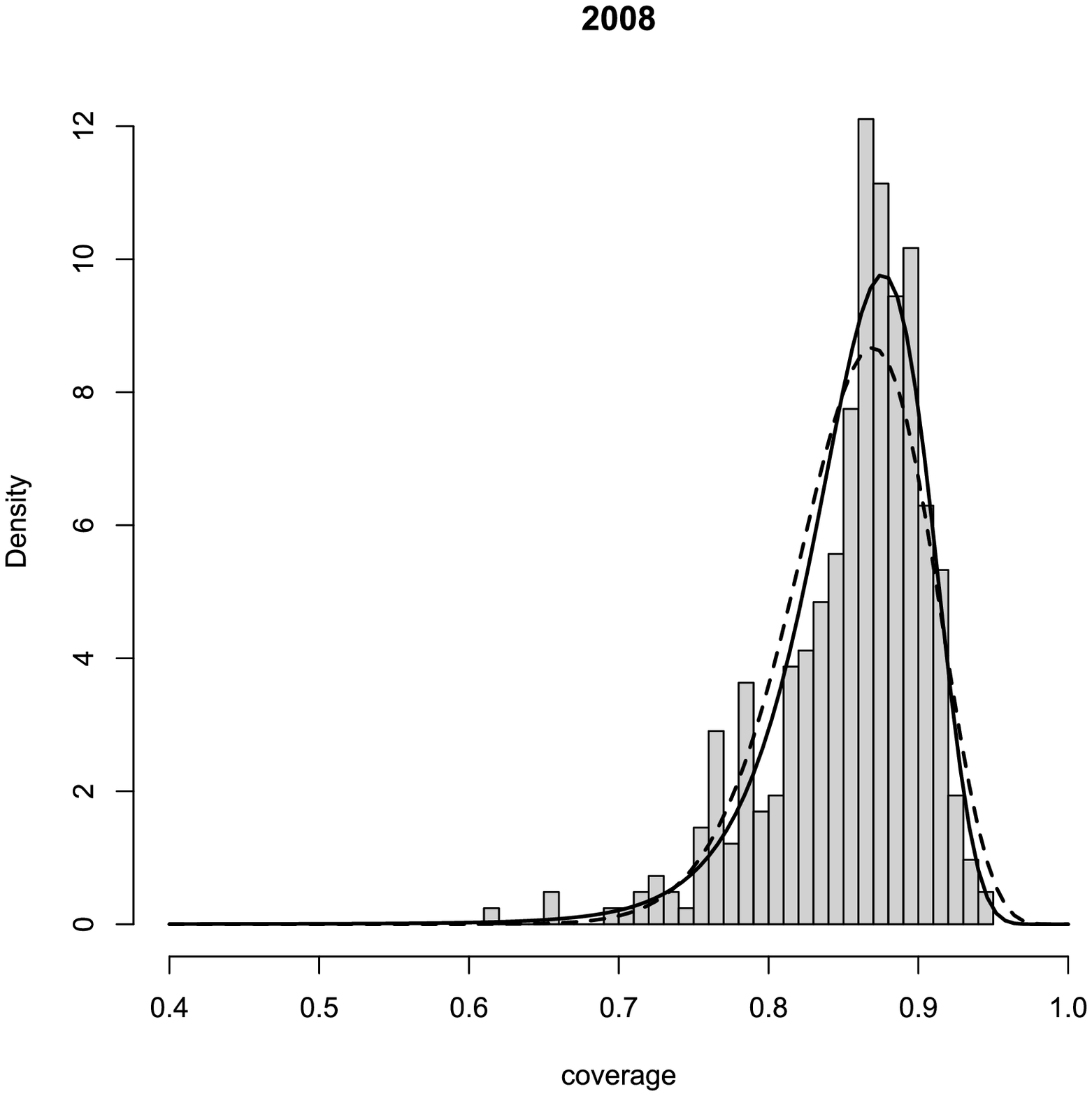}
		\includegraphics[width=7cm]{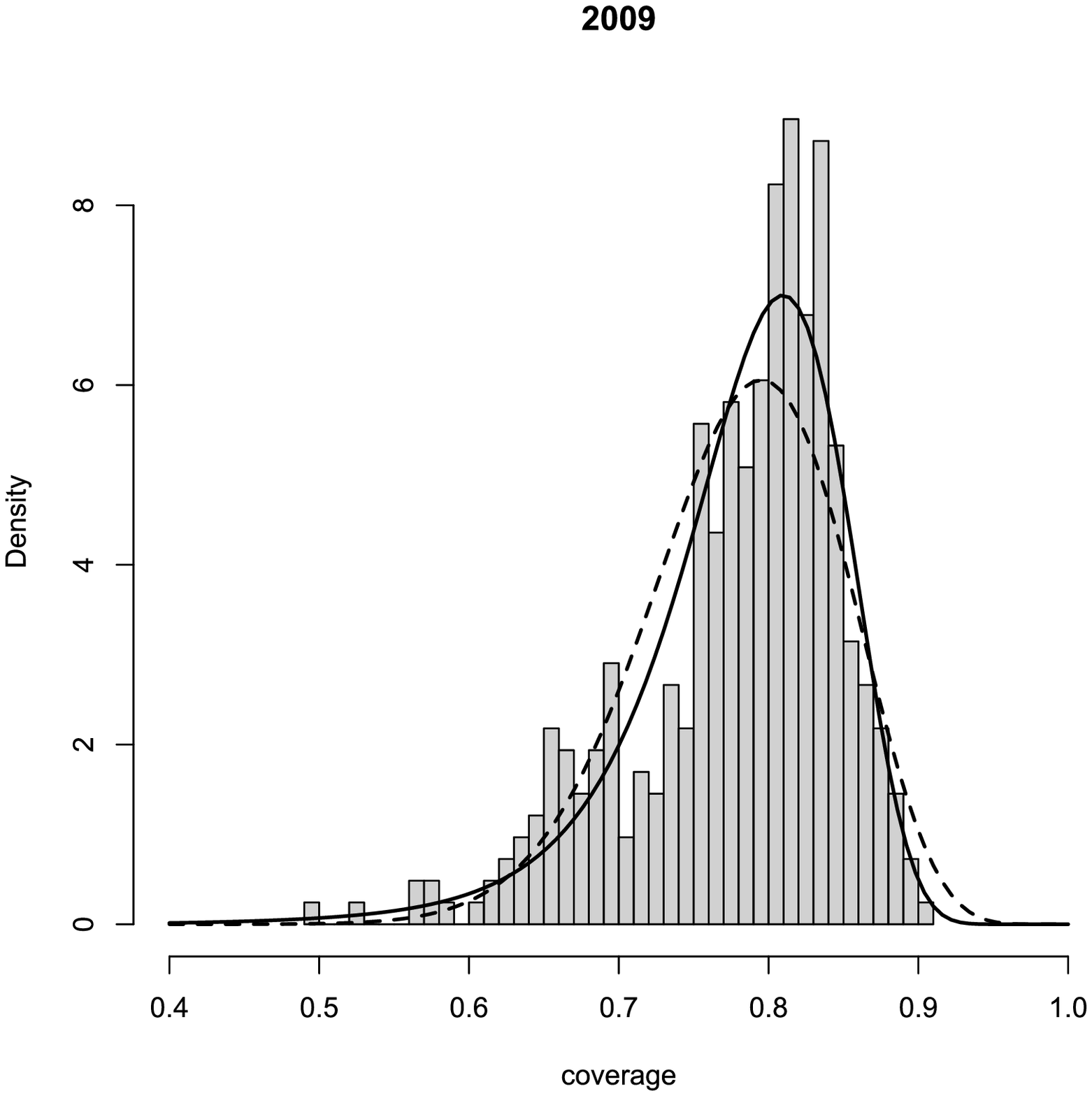}
		\includegraphics[width=7cm]{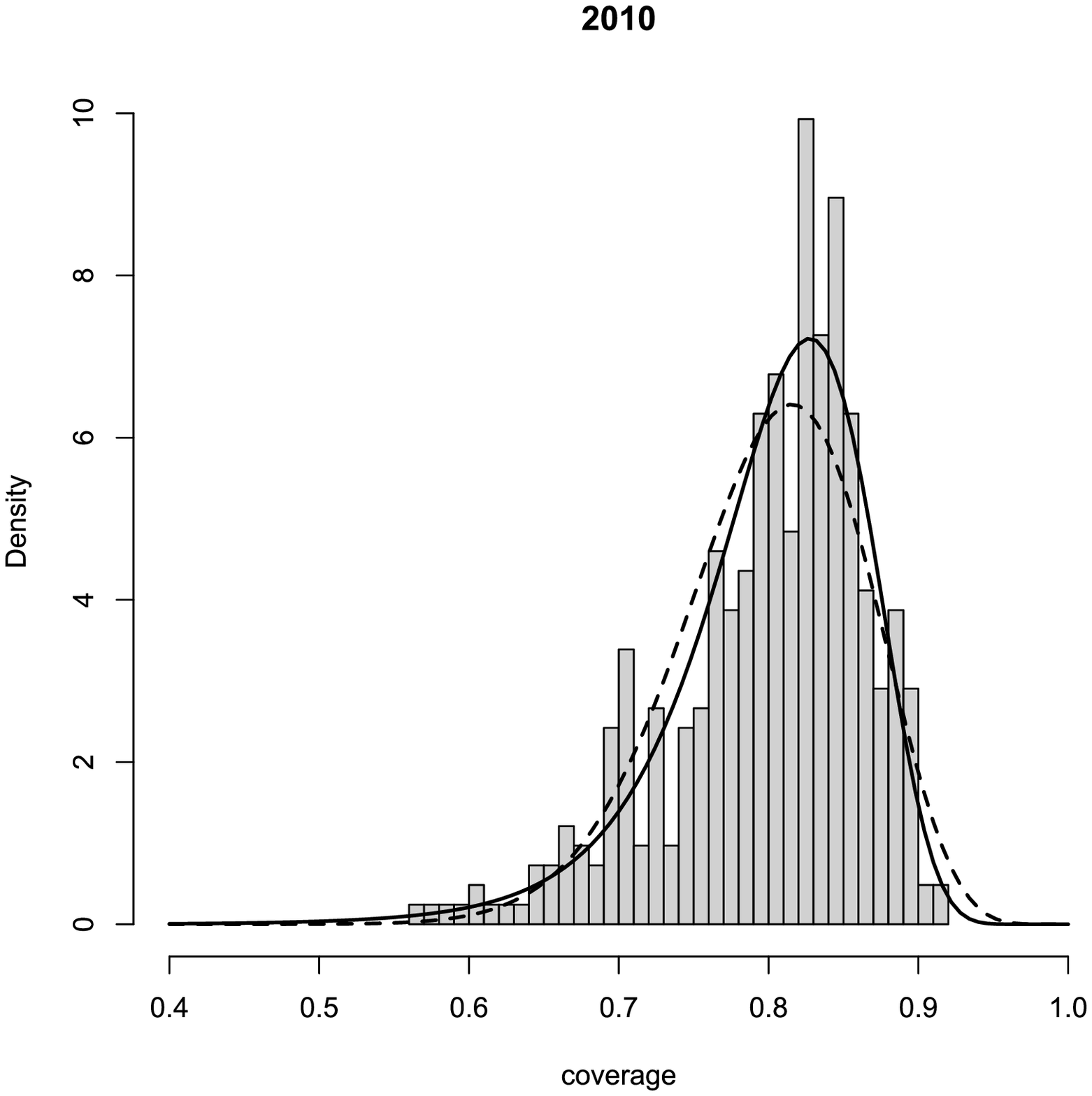}
		\includegraphics[width=7cm]{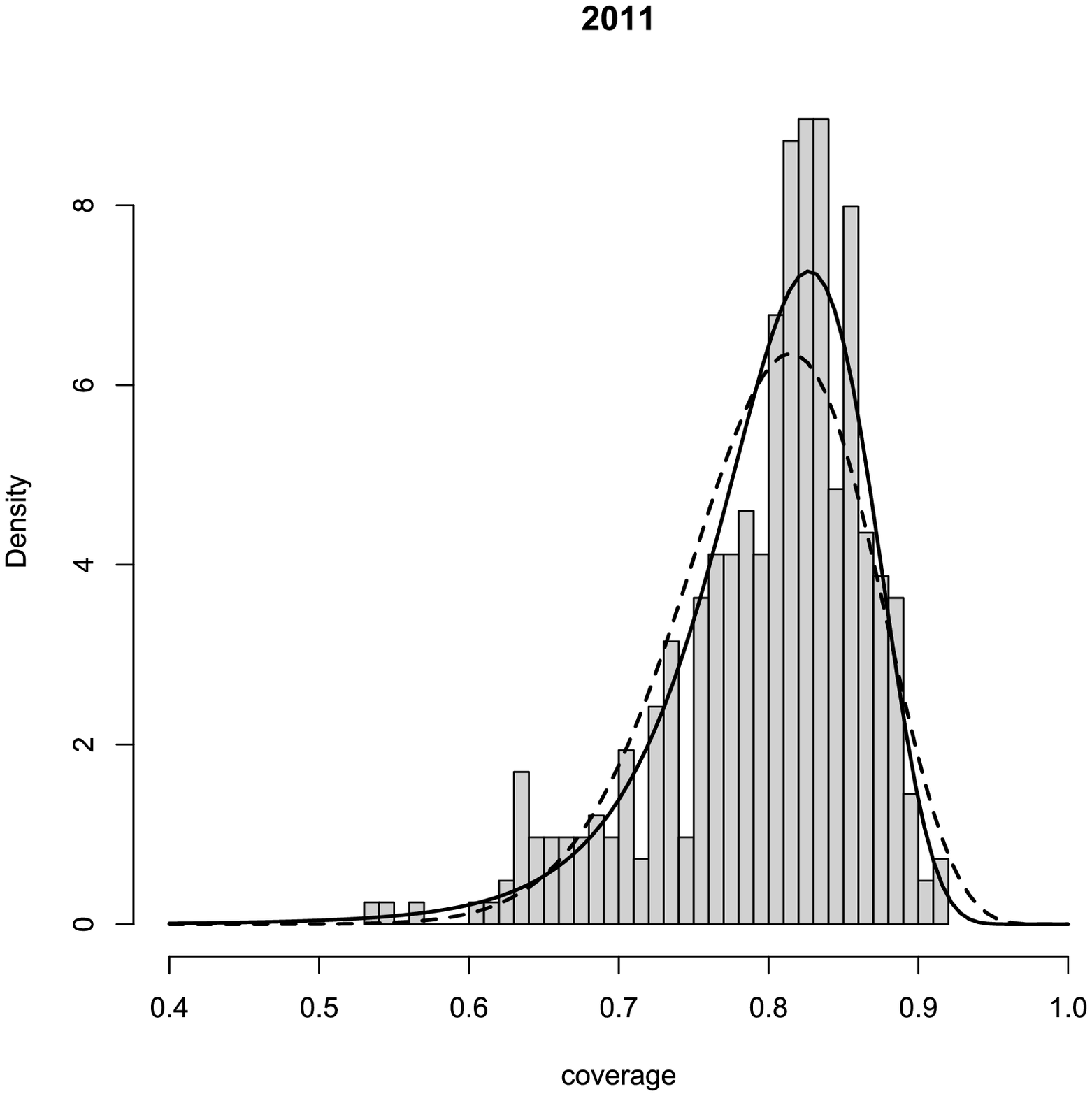}
		\includegraphics[width=7cm]{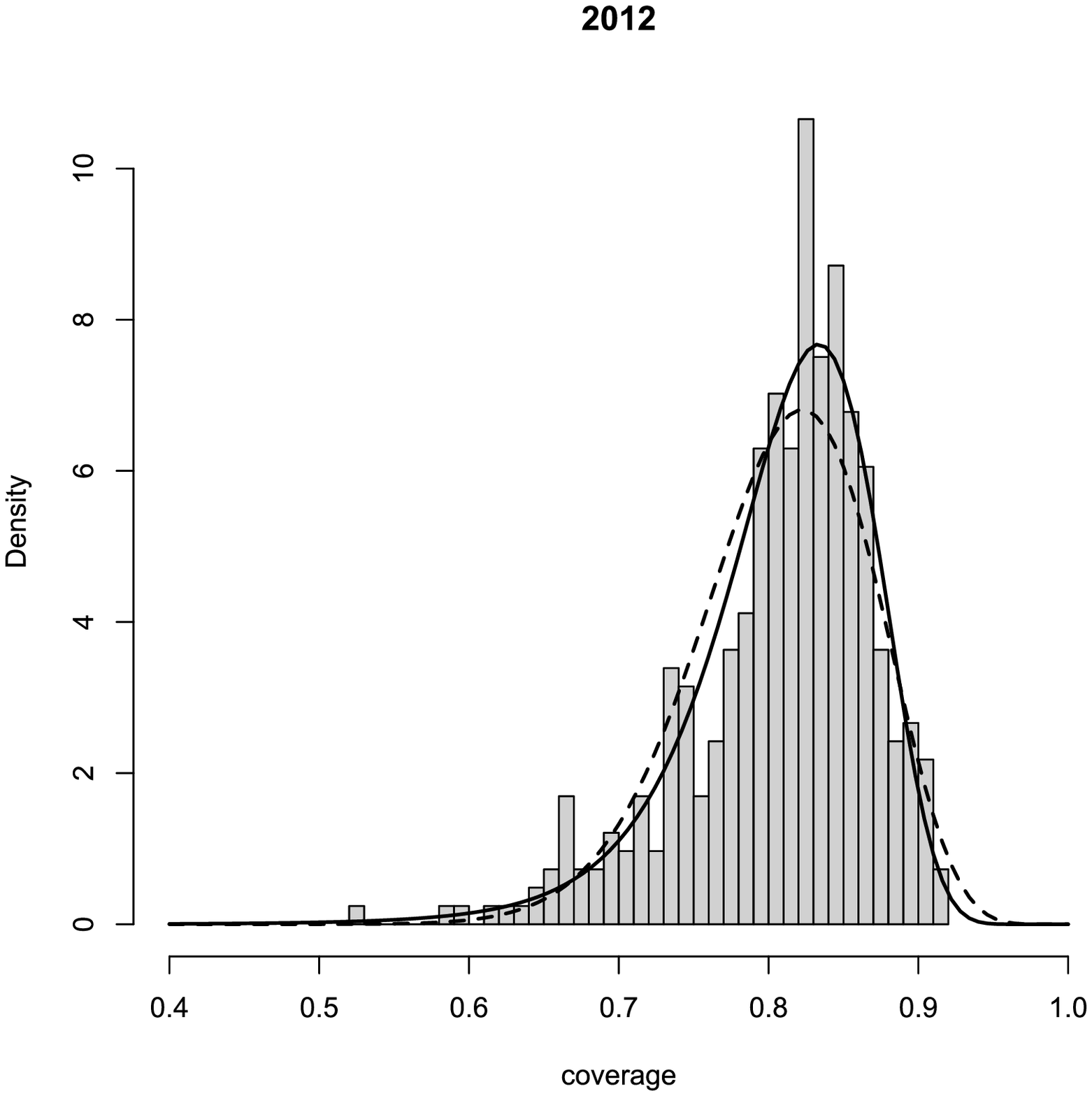}
	\end{center}
	\caption{Histograms for the measles vaccination coverage (decoupled model). Solid line: Probability density with reinforcement, dashed line: probability density without reinforcement.}
\end{figure}

\begin{figure}
	\begin{center}
		\includegraphics[width=7cm]{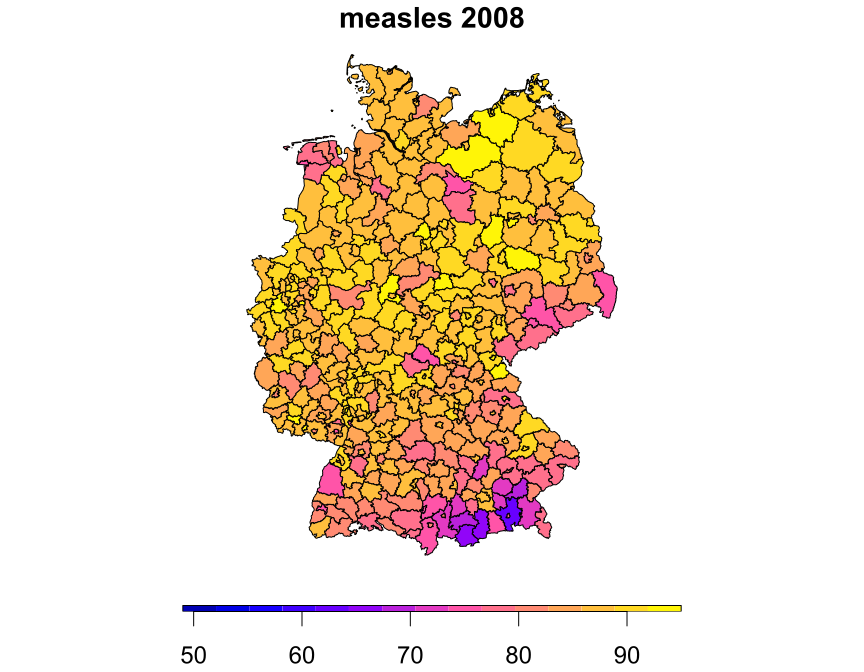}
		\includegraphics[width=7cm]{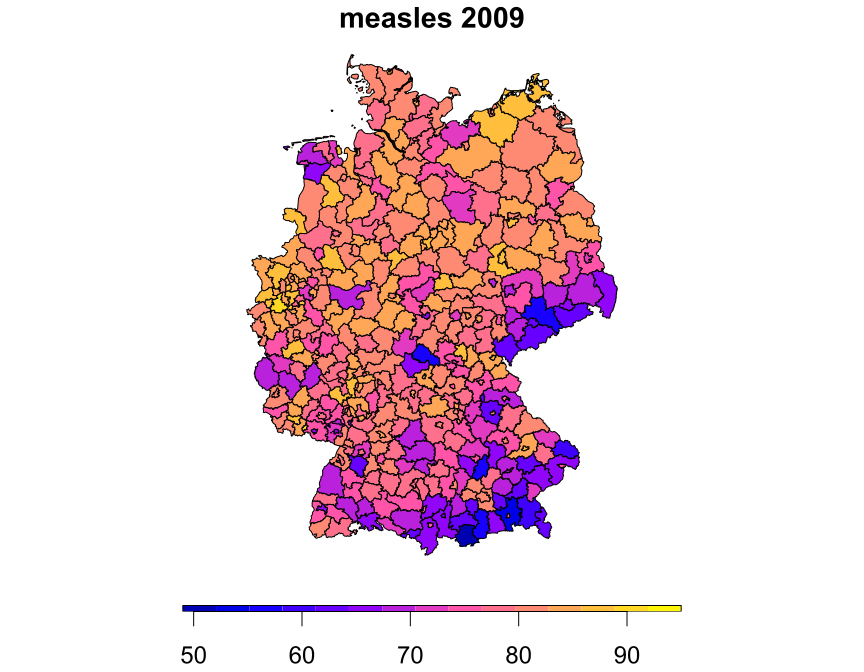}
		\includegraphics[width=7cm]{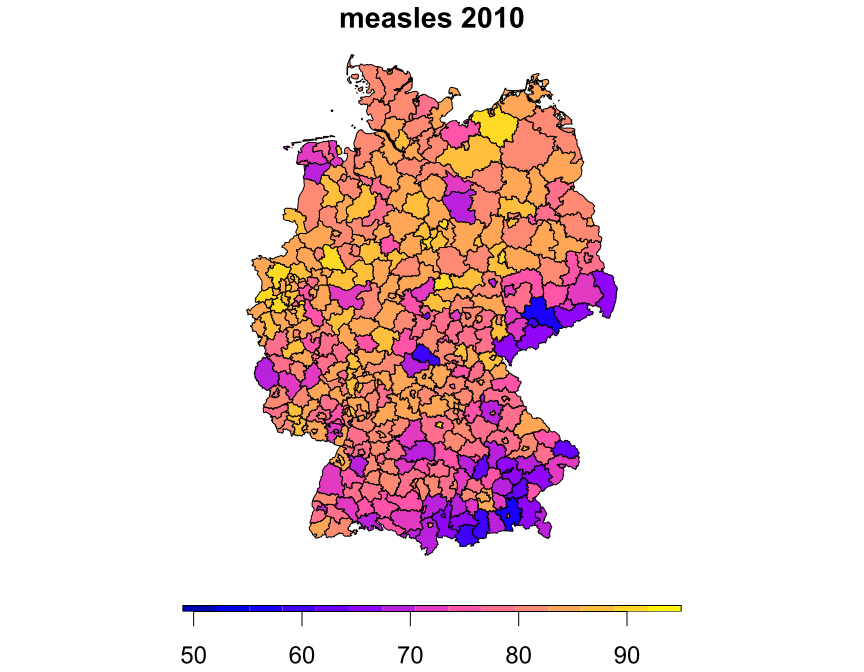}
		\includegraphics[width=7cm]{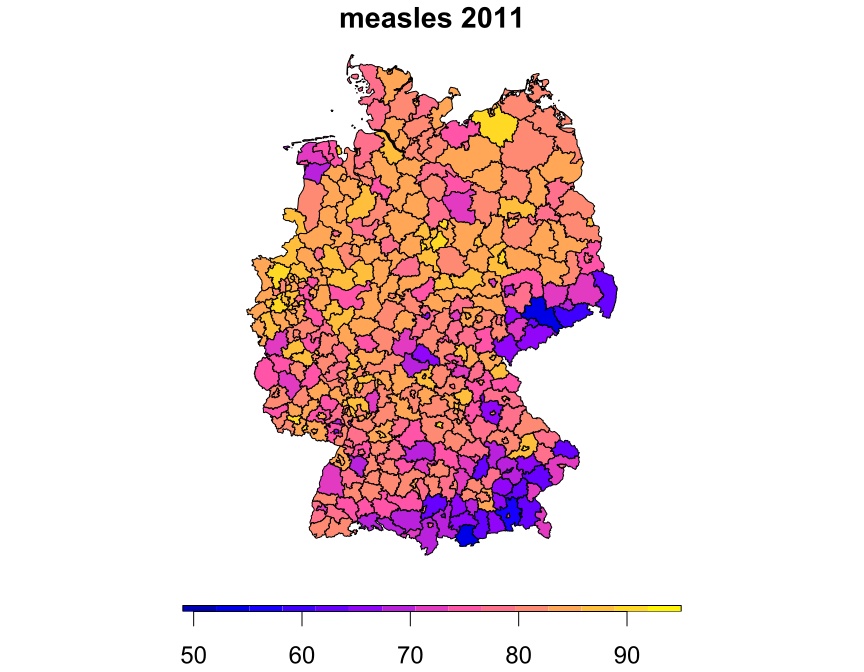}
		\includegraphics[width=7cm]{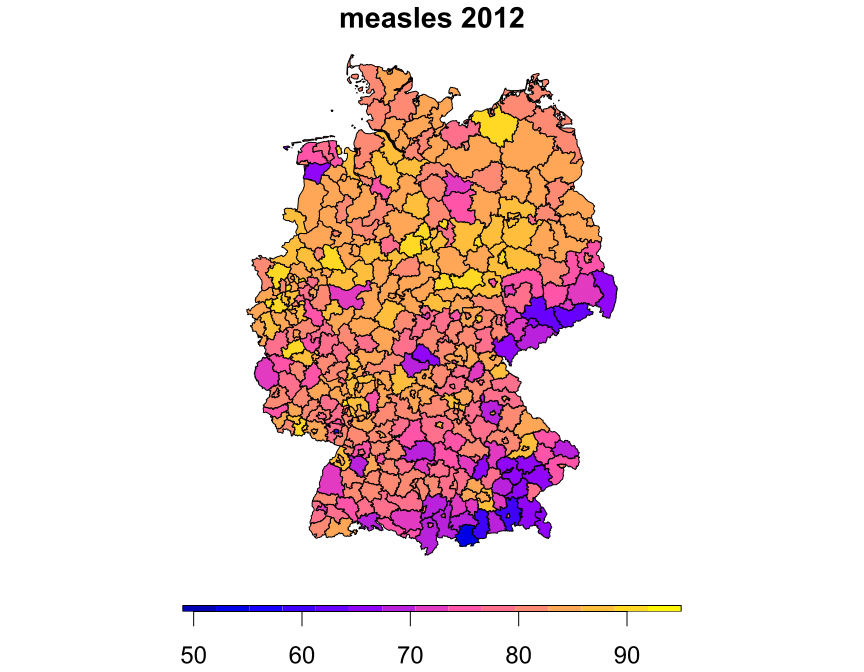}
	\end{center}
	\caption{Vaccination coverage (in percent) for measles in Germany.}\label{measlesGIS}
\end{figure}

\begin{figure}[h!]
	\begin{center}
		\includegraphics[width=7cm]{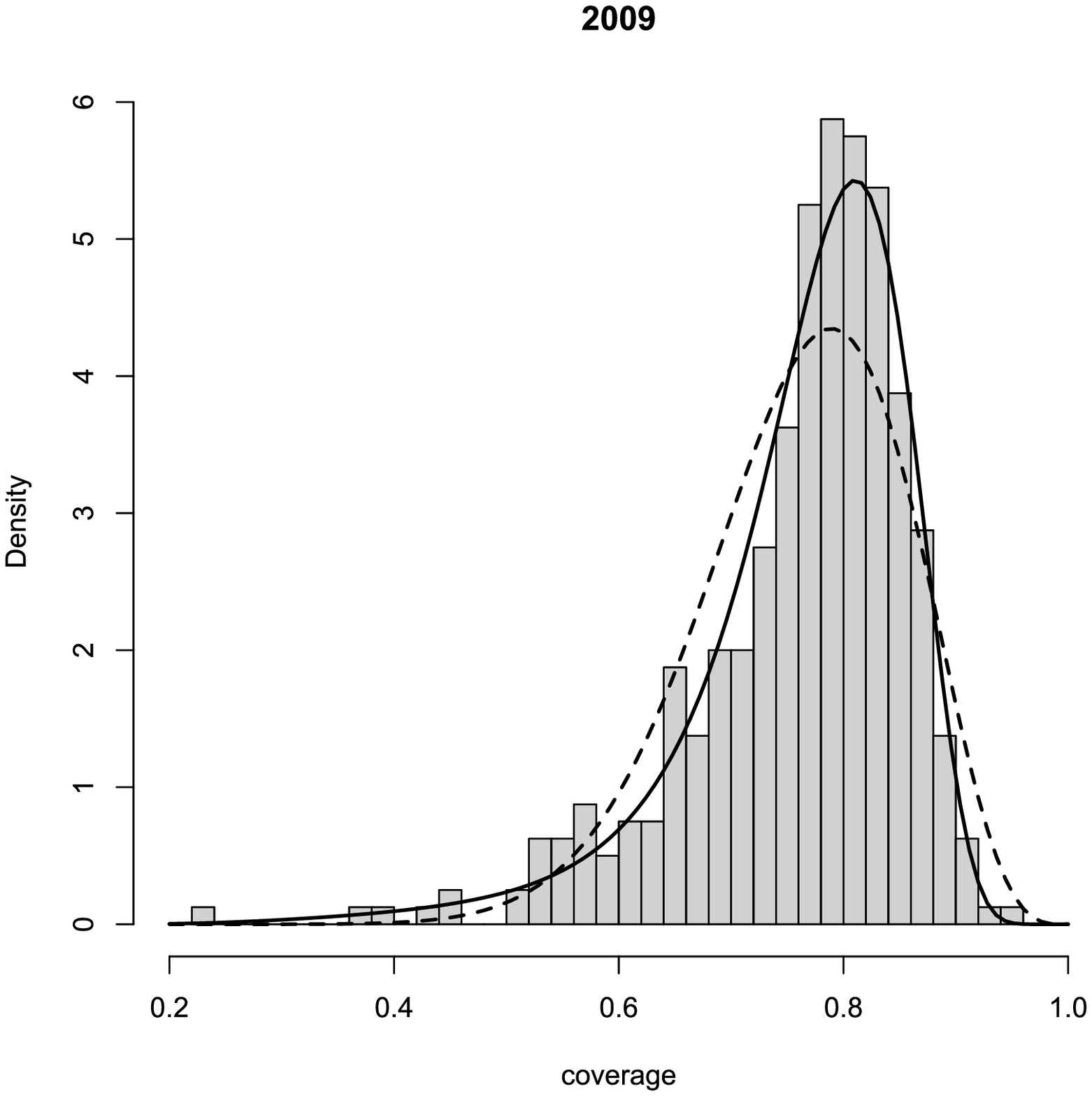}
		\includegraphics[width=7cm]{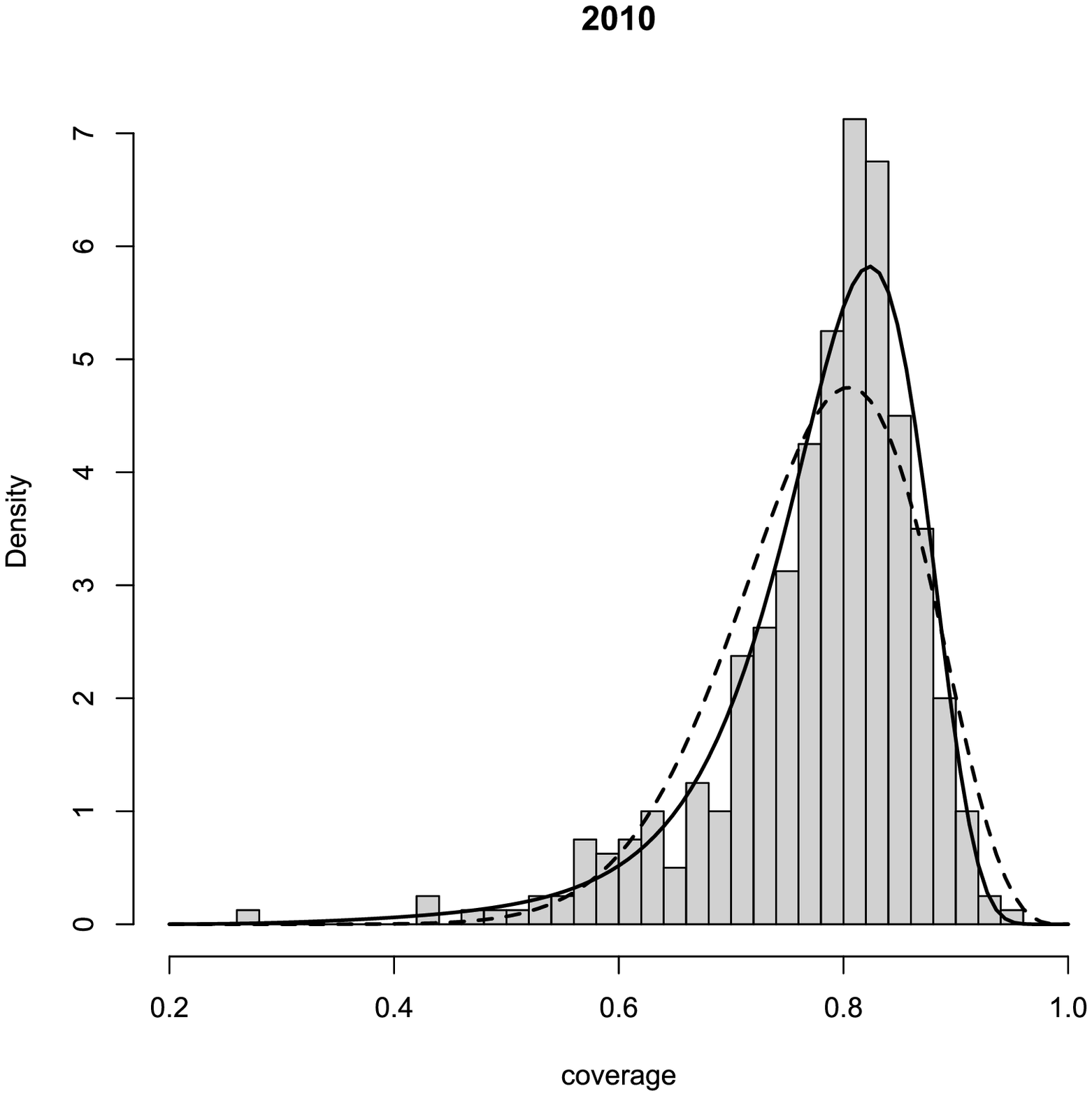}
		\includegraphics[width=7cm]{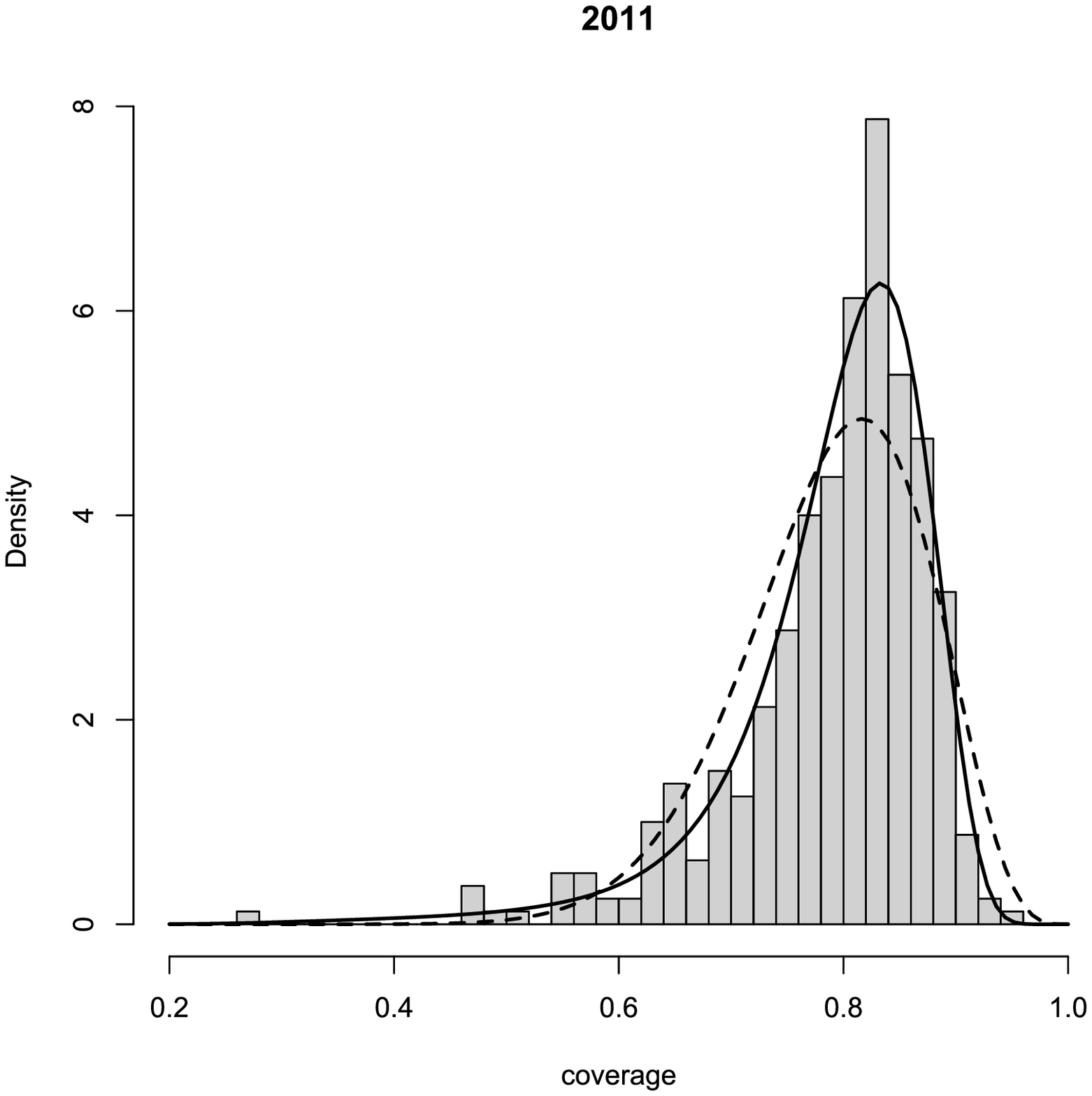}
		\includegraphics[width=7cm]{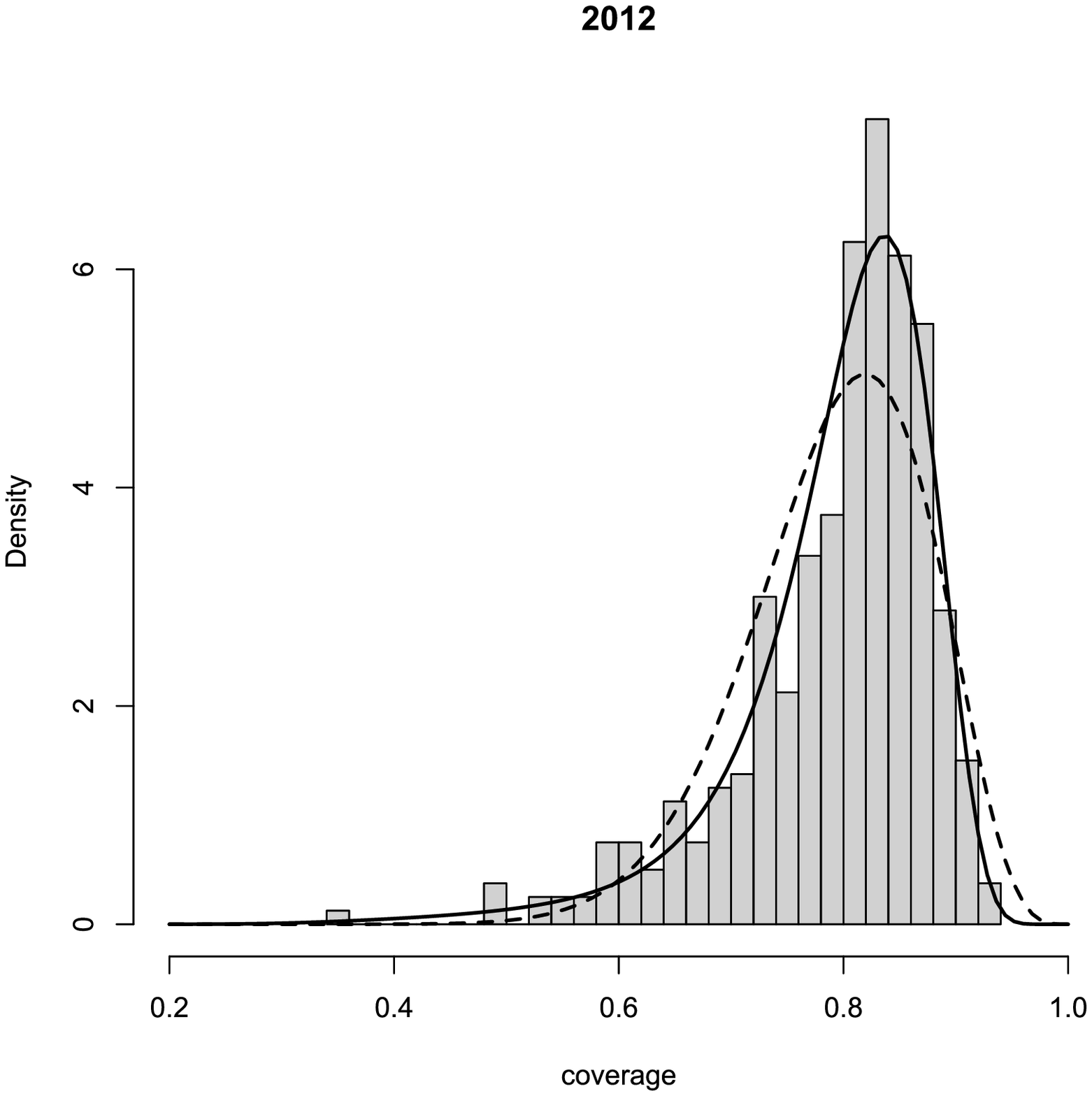}
		\includegraphics[width=7cm]{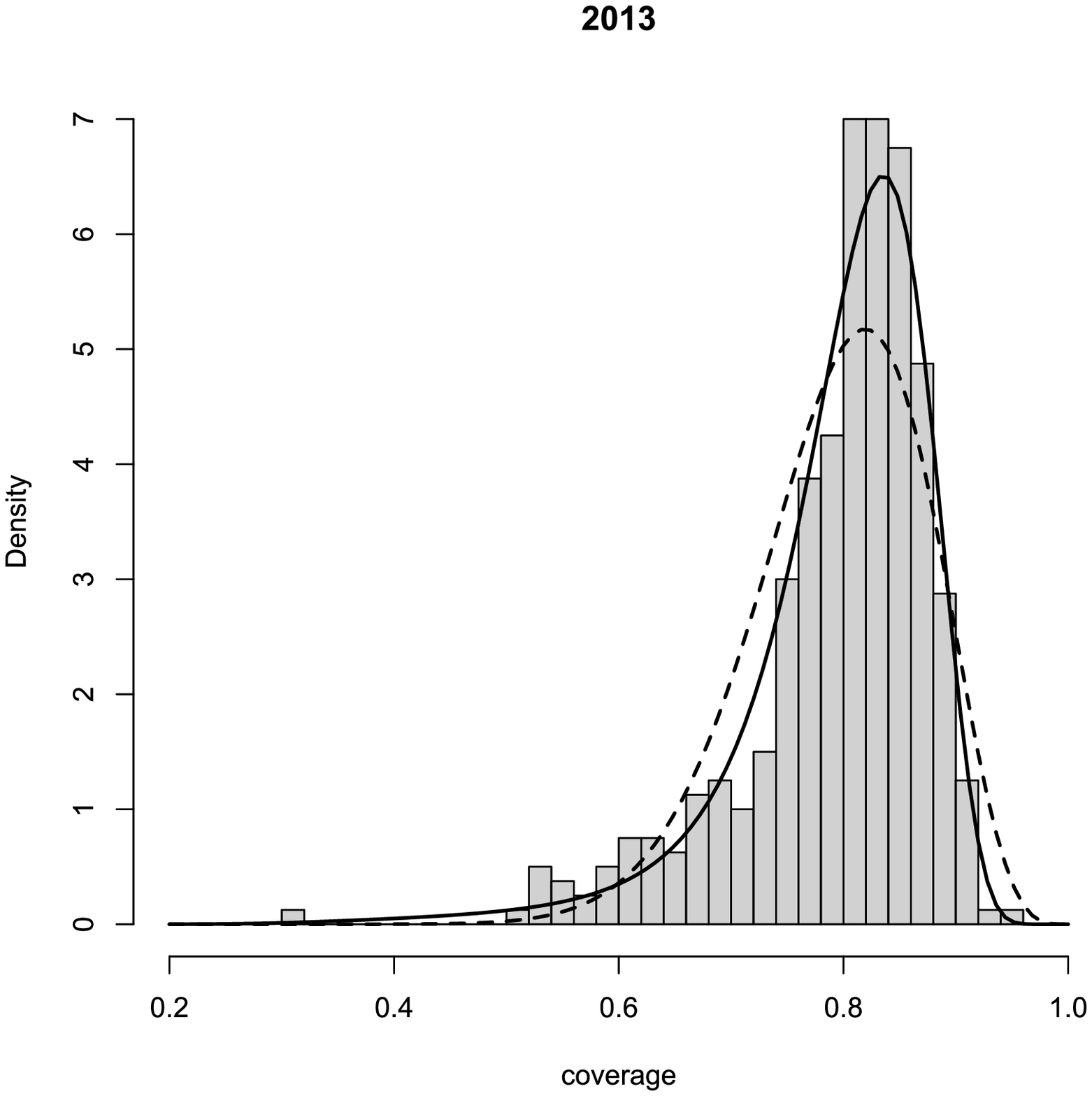}
	\end{center}
	\caption{Histograms for the meningicocci vaccination coverage (homogeneous model). Solid line: Probability density with reinforcement, dashed line: probability density without reinforcement.}
\end{figure}

\begin{figure}
	\begin{center}
		\includegraphics[width=7cm]{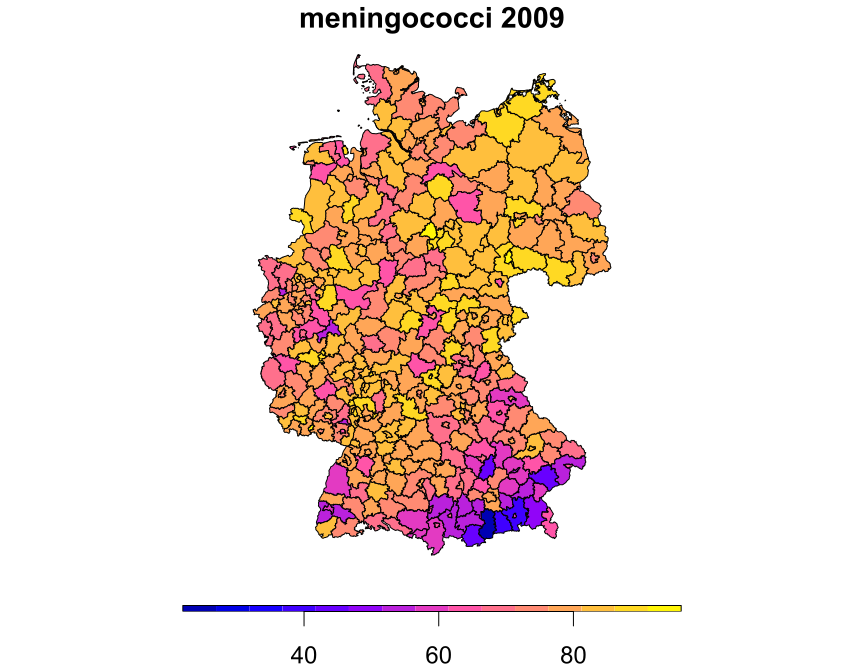}
		\includegraphics[width=7cm]{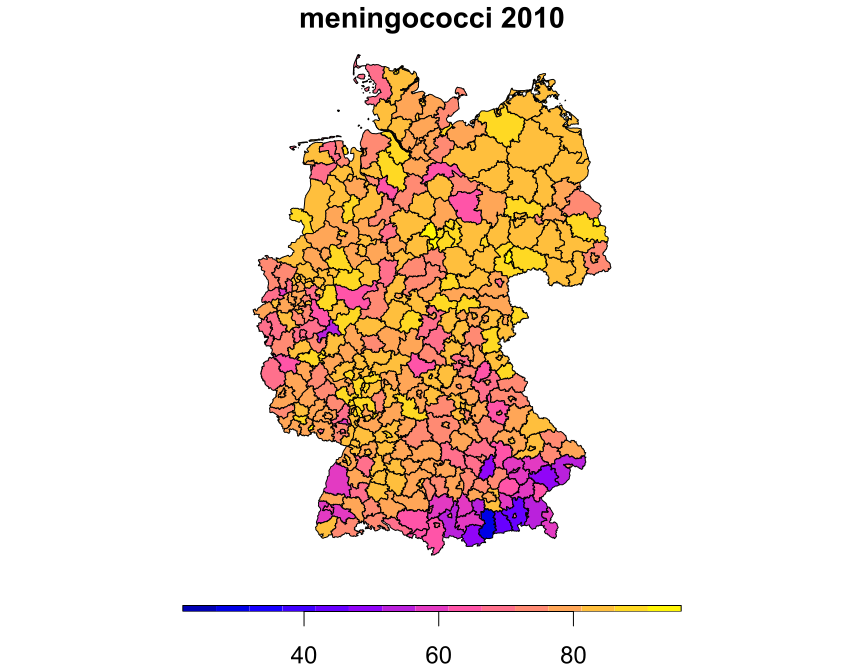}
		\includegraphics[width=7cm]{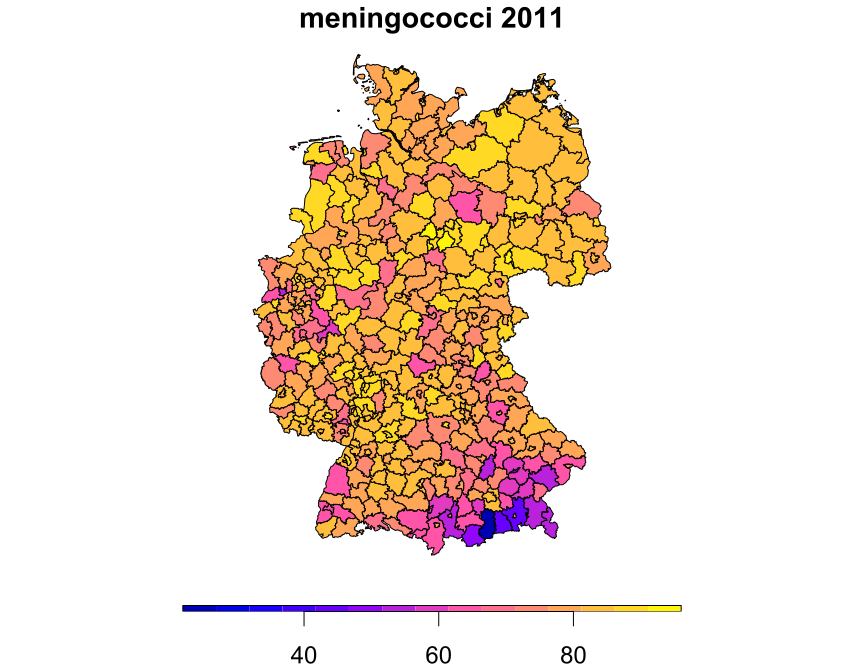}
		\includegraphics[width=7cm]{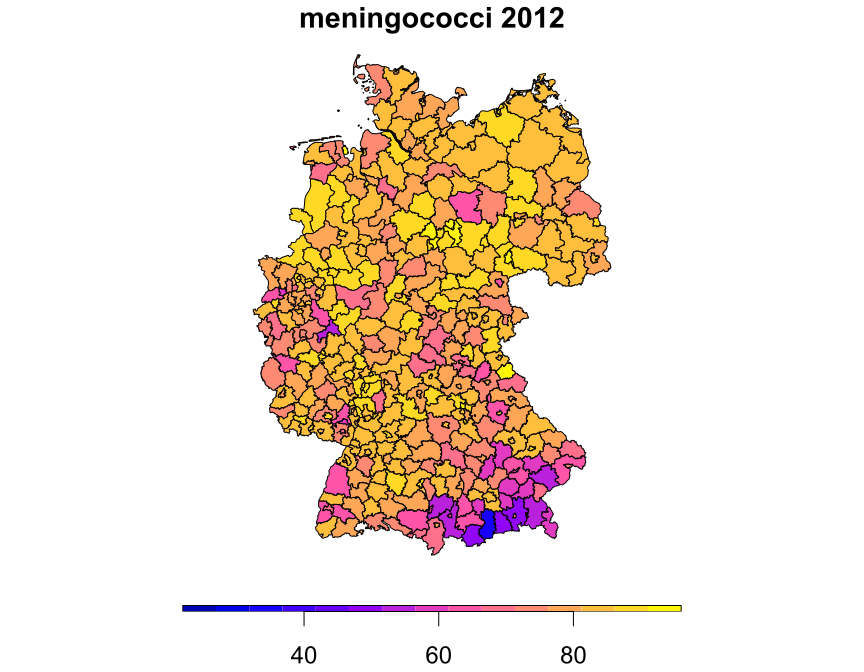}
		\includegraphics[width=7cm]{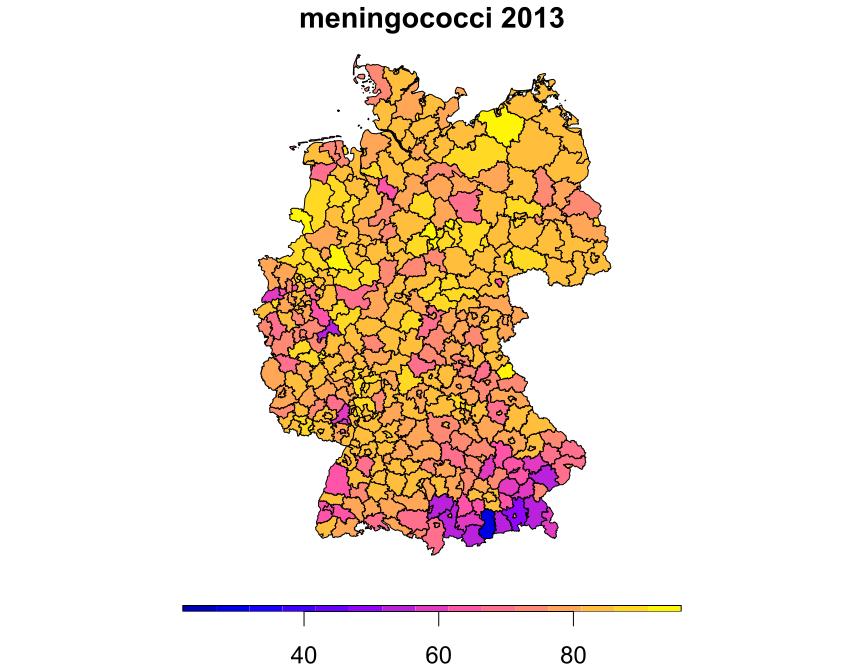}
	\end{center}
	\caption{Vaccination coverage (in percent) for meningococci in Germany. Note that the region of Saxonia is missing as its official recommendations for the meningococci vaccination differ from the rest of Germany.}\label{meningoGIS}
\end{figure}

\end{appendix}

\end{document}

%% file: datAnaMeaslesHomoV1.tex
 2008  & 59.64  &  15.50  & 250.71  &  92.65  & 6.88e-07  &  0.072  &  0.0011  & 0.86 \\
 2009  & 57.47  &  22.38  & 213.05  &  102.61  & 4.84e-08  &  0.24  &  0.0038  & 0.78 \\
  2010  & 53.14  &  17.97  & 188.38  &  82.01  & 4.11e-06  &  0.40  &  0.019  & 0.80 \\
 2011  & 47.32  &  18.49  & 177.11  &  91.58  & 2.30e-07  &  0.23  &  0.0018  & 0.80 \\
 2012  & 50.91  &  18.56  & 189.53  &  92.49  & 2.07e-06  &  0.57  &  0.014  & 0.81 \\

%% file: datAnaMeaslesSpatV1.tex
  2008  & 45.29  &  11.90  & 317.61  &  90.70  & 1024.70 \\
 2009  & 49.49  &  19.86  & 271.88  &  116.27  & 540.52 \\
 2010  & 41.33  &  14.35  & 237.79  &  88.04  & 593.66 \\
 2011  & 47.14  &  15.72  & 259.09  &  93.55  & 556.15 \\
  2012  & 38.42  &  14.15  & 243.34 &  93.59  & 700.49 \\

%% file: datAnaMeaslesSpatNoSaxV1.tex
  2008  & 46.72  &  12.46  & 328.42  &  95.90  & 1019.21 \\
 2009  & 53.81  &  21.56  & 294.87  &  126.26  & 522.32 \\
 2010  & 44.01  &  15.26  & 250.89  &  93.33  & 574.34 \\
  2011  & 48.73  &  16.43  & 269.03  &  98.30  & 549.74 \\
  2012  & 40.73  &  15.20  & 258.63  &  100.94  & 690.08 \\

%% file: datAnaMeningoHomoV1.tex
 2009  & 22.11 &  13.76  & 84.82  &  70.80  & 8.66e-12  &  0.75  &  0.0013  & 0.76 \\
 2010  & 28.68  &  13.45  & 108.34  &  68.03  & 1.25e-10  &  0.57  &  0.0028  & 0.78 \\
  2011  & 29.19  &  13.89  & 119.42  &  75.99  & 2.62e-14  &  0.46  &  0.00069  & 0.79 \\
  2012  & 41.39  &  13.93  & 159.23  &  67.96  & 2.56e-12  &  0.57  &  0.0013  & 0.80 \\
  2013  & 39.58  &  14.79  & 157.14  &  75.99  & 2.14e-13  &  0.56  &  0.00090  & 0.80 \\

%% file: datAnaMeningoSpatNoSaxV1.tex
 2009  & 16.83  &  10.56  & 91.10  &  62.012  & 248.97 \\
 2010  & 22.39  &  10.66  & 116.66  &  62.77  & 255.67 \\
 2011  & 26.19  &  11.87  & 144.67  &  73.77  & 277.62 \\
 2012  & 48.82  &  12.55  & 223.89  &  63.24  & 286.69 \\
  2013  & 39.54  &  12.52  & 192.78  &  69.66  & 273.14 \\

%% file: elast.tex
Measles  &  2008  &  0.21  &  -0.28  &  -0.133  &  0.22  \\
 Measles  &  2009  &  0.48  &  -0.58  &  -0.34  &  0.46  \\
 Measles  &  2010  &  0.38  &  -0.45  &  -0.24  &  0.33  \\
 Measles  &  2011  &  0.35  &  -0.47  &  -0.24  &  0.38  \\
 Measles  &  2012  &  0.32  &  -0.42  &  -0.21  &  0.34  \\
 \hline
 Meningo  &  2009  &  0.41  &  -0.64  &  -0.32  &  0.60  \\
 Meningo  &  2010  &  0.42  &  -0.56  &  -0.32  &  0.50  \\
 Meningo  &  2011  &  0.41  &  -0.56  &  -0.33  &  0.53  \\
 Meningo  &  2012  &  0.54  &  -0.55  &  -0.40 &  0.46  \\
 Meningo  &  2013  &  0.49  &  -0.56  &  -0.38  &  0.48  \\